CHEMICAL SENSING WITH NANOGAP AND CHEMI-MECHANICAL DEVICES

by

Aishwaryadev Banerjee

A dissertation submitted to the faculty of
The University of Utah
in partial fulfillment of the requirements for the degree of

Doctor of Philosophy

Department of Electrical and Computer Engineering

The University of Utah

December 2019

**The University of Utah Graduate School**

**STATEMENT OF DISSERTATION APPROVAL**

The          **Aishwaryadev Banerjee** ___________________

dissertation of has been approved by the following supervisory committee members:

| | | |
|---|---|---|
| **Carlos H. Mastrangelo** , | Chair | 04/23/19 |
| | | Date Approved |
| **Massood Tabib-Azar** , | Member | 04/23/19 |
| | | Date Approved |
| **Hanseup Kim** , | Member | 04/23/19 |
| | | Date Approved |
| **Florian Solzbacher** , | Member | 06/06/19 |
| | | Date Approved |
| **Shadrach J Roundy** , | Member | 04/23/19 |
| | | Date Approved |

and by ________ **Florian Solzbacher** ________ , Chair/Dean of the

Department/College/School of ________ **Electrical and Computer Engineering** ________

and by David B. Kieda, Dean of The Graduate School.

## ABSTRACT

This dissertation presents a new class of batch-fabricated, low-power and highly sensitive chemiresistive sensors. We first present the design, fabrication, and characterization of batch-fabricated sidewall etched vertical nanogap tunneling-junctions for bio-sensing. The device consists of two vertically stacked gold electrodes separated by a partially etched sacrificial spacer-layer of $\alpha$-Si and $SiO_2$. A ~10 nm wide air-gap is formed along the sidewall by a controlled dry etch of the spacer, whose thickness is varied from ~$4.0 - 9.0$ nm.

Using these devices, we demonstrate the electrical detection of certain organic molecules from measurements of tunneling characteristics of target-mediated molecular junctions formed across nanogaps. When the exposed gold surface in the nanogap device is functionalized with a self-assembled monolayer (SAM) of thiol linker-molecules and then exposed to a target, the SAM layer electrostatically captures the target gas molecules, thereby forming an electrically conductive molecular bridge across the nanogap and reducing junction resistance.

We then present the design, fabrication and response of a humidity sensor based on electrical tunneling through temperature-stabilized nanometer gaps. The sensor consists of two stacked metal electrodes separated by ~2.5 nm of a vertical air gap. Upper and lower electrodes rest on separate 1.5 μm thick polyimide patches. When exposed to a humidity change, the patch under the bottom electrode swells but the patch under the top electrode does not, and the air gap thus decreases leading to increase in the tunneling current across the junction.

Finally, we present an electrostatic MEMS switch which is triggered by a very low input voltage in the range of ~50mV. This consists of an electrically conductive torsional see-saw paddle with four balanced electrodes. It is symmetrically biased by applying the same voltage at its inner electrodes leading to bistable behavior at flat or collapsed equilibrium positions. The use of elevated symmetric bias softens the springs such that the paddle collapses when a few milliVolts are applied to one of its outer electrodes thus causing the device to snap in and result in switch closure. Using the "spring softening" principle, we also present an application of a new kind of high sensitivity chemo-mechanical sensors.

To The Struggling Scholar

Q: "By when do I have to get the results?  A: Yesterday"

<div align="right">– C.H.M</div>

# TABLE OF CONTENTS

# CHAPTER 1

# INTRODUCTION

## 1.1 <u>Motivation</u>

In today's world of ever-increasing power-hungry applications, inexpensive, portable devices which consume low power are crucial in building a truly interconnected and 'smart' civilization. With the advent of Internet of Things (IoT), we seem to be inching closer towards realizing such a society. The IoT is a concept which describes an interconnected system of physical devices which communicate with each other over the internet and can be remotely monitored and controlled [1]. The first internet connected appliance was a Coke vending machine at Carnegie Mellon University that could report if the drinks loaded inside the machine were cold or not. Since then, devices intended to be part of the IoT have evolved greatly. The number of devices which constitute the IoT network exceeded the world population way back in 2008. It is projected that this number will reach 50 billion in the year 2020 and the IoT market will surpass the market of the PC, tablet and phone combined. The potential growth in this industry is extremely high since only 0.06% of all possible devices have been optimized for IoT [2]. These devices include consumer electronics such as 'smart thermostats' [3] which allow consumers to remotely control the temperature settings of their house as well as sophisticated MEMS sensors deployed in vehicles which provide driving assistance, optimized logistics and predictive maintenance [4]. Since these devices have to be remotely present across the IoT network grid, it is critical that they fulfill certain requirements. First and most importantly, they need to consume low power. Many IoT applications involve remote air quality monitoring and asset tracking. Such applications require the devices to be battery operated and deployed to remote areas of the world. Therefore, low power consumption is of utmost importance. Second, these devices also need to be highly portable. Consumer-electronic products such as the 'smart-watch' or applications such as remote-monitoring of pacemaker implants require the communicating devices to be as small as possible. Hence, large and bulky devices are not suitable for such applications. Third, these devices need to provide accurate and legitimate information for precise data collection and further analysis. For example in remote sensing and medical applications, the employed devices cannot be susceptible to faulty data. Since most of these devices are no longer implemented as part of 'stand-alone' functionalities, failure of one device will inevitably lead to a cascading failure of another. This domino effect needs to be prevented at all costs.

Keeping these certain requirements in mind, in the following sections we discuss the various aspects of modern-day gas sensor technology and analyze why most of them cannot be used for IoT purposes. We also provide a brief discussion of other sensor technologies and sensor systems which have been developed specifically for use in the IoT framework. These devices and their major drawbacks are summarized in the following sections.

## 1.2 Modern Sensor Technologies

Existing sensor technology can be divided into mainly five categories: conductivity based, solid-state, optical, piezoelectric and polymer swelling induced gas sensors.

### 1.2.1 Conductivity sensors

In these types of sensors, the conductivity of the analyte-sensitive material changes when exposed to the analyte. Most common realizations of these devices include the use of conducting polymer composites [5], conducting polymers [6, 7] and metal oxides [7] as the sensing material. When the sensing layer comprised of a conductive polymer composite such as PEDOT:PSS or polypyrrole is exposed to the analyte, the polymer film absorbs the vapor which causes it to swell. This expansion causes a reduction in current conduction paths along the polymer which leads to an increase in resistance of the polymer film. Intrinsically conductive polymers such as polyaniline are also used as gas sensitive materials for such sensors. The principle of operation is the same as described above: exposure to a gaseous analyte leads to an expansion of the polymer which causes a change in electron density of the polymer chains and this changes the resistance of the polymer itself. Metal-oxide (MOX) based gas sensors work on the principle that the oxide (either p-type which respond to oxidizing gases or n-type which respond to reducing gases) reacts with the appropriate gas which leads to an excess of majority charge carriers (holes or electrons in p-type or n-type sensors). These excess charge carriers lead to augmented current conduction. Commonly used MOX sensors employ tin dioxide, zinc oxide, nickel oxide, cobalt oxide and iron (III) oxide as part of their sensing layer.

These devices have certain inherent advantages such as conductive polymer based sensors display reasonable selectivity in sensor response, they are relatively cheap to prepare and show linear response for a wide range of target analytes. MOX sensors show fast response and recovery times. However, the major disadvantage with these sensors is that they require high operating temperatures to function properly and hence cannot be used for low power applications. Figure 1.1 shows a schematic of conductive gas sensor devices and their working principle.

### 1.2.2 Solid-state gas sensors

In these sensors, the threshold voltage of the semiconductor device [8, 9] (typically a MOSFET or PolFET) changes when exposed to a target analyte. This is because the interface of the catalytic metal (gate electrode) and the oxide layer gets polarized when exposed to the gaseous analyte and this changes the work-function of the metal and oxide layer. To facilitate a reaction between the analyte and the metal-insulator interface, usually a porous gas sensitive gate metal, such as Pd or a suspended gate design is used to provide access to the metal-insulator interface. These devices can be microfabricated using standard CMOS techniques which makes them highly compatible with existing CMOS circuitry and they are cheap to manufacture. However, the major drawbacks of these types of sensors is that they suffer from baseline drift and instability. They require complicated packaging and can function well only if the surrounding environment is controlled. Hence, they cannot be deployed for remote sensing purposes. Figure 1.2 shows a schematic representation of the working principle of solid-state gas sensors.

### 1.2.3 Optical gas sensors

In optical gas sensors, an optical fiber is coated with a fluorescent dye such as Pyranine or HPTS. This is encapsulated within a polymer matrix. When the optical fiber interacts with the target gas, the optical properties of the dye such as intensity, spectrum or wavelength change [10]. The sensitivity of the sensor depends on the type of dye and the polymer in which the dye is imbedded. Typically, adsorbents such as $Al_2O_3$ are often added to the polymer for improving detection limits [11]. The major advantages of these devices are that they have extremely fast response times and are immune to electromagnetic interference. However, to use them, one needs to implement complicated electronics and postprocessing software algorithms which makes them impractical for IoT based applications. Additionally, these sensors are also susceptible to photobleaching which can render them ineffective. Therefore, these sensors cannot be a part of the IoT framework. Figure 1.3 shows the schematic of an optical gas sensor and its working principle.

### 1.2.4 Piezoelectric gas sensors

In piezoelectric gas sensors, the resonant frequency of the output signal changes when exposed to the target analyte. A typical surface acoustic wave (SAW) device includes two interdigitated transducers on a piezoelectric substrate (e.g. ZnO or Lithium niobate) with a polymeric gas sensitive coating in between [12]. An AC signal applied at the input electrode produces two-dimensional waves which travel along the surface of the substrate. When the device is exposed to the analyte, the coating adsorbs the analyte molecules and this change in mass of the gas sensitive coating changes the frequency of the traveling surface wave, which is then sensed at the output electrode. Similar to the SAW devices, quartz crystal microbalance (QCM) devices [13] operate on the principle that exposure to certain analytes leads to an increase in the mass of the gas sensitive material, which is deposited on a quartz crystal. This change in mass leads to a change in resonant frequency of the crystal and is used for detecting presence of analyte. The advantages of these devices are that they offer high sensitivity and fast response times. However, they require complex and expensive circuitry to function. Additionally, batch to batch reproducibility and low SNR add to the list of disadvantages of these devices. Figure 1.4 shows a typical piezoelectric gas sensor device.

### 1.2.5 Polymer swelling induced gas sensors

MEMS cantilevers have been widely used as gas sensors for more than a decade. In a typical MEMS-based humidity sensor [14], a Si free-standing cantilever is coated with a patch of polyimide, which is a moisture-sensitive polymer. The cantilever is suspended at a small distance on top of another electrode. When the device is exposed to change in ambient moisture, the polymer swells which causes a differential strain in the polymer-Si bimorph cantilever. This causes the cantilever to bend, thereby reducing the distance between the top cantilever and bottom electrode. The change in this distance causes the capacitance to increase which is a measure of the absorbed humidity. Such sensors have been widely used since they typically consume very low power and can be batch-fabricated. However, they are extremely temperature sensitive and display a lack of selectivity amongst different analytes [15]. Pattern recognition algorithms have been used to significantly improve the selectivity of these sensors. Figure 1.5 shows the working of MEMS based cantilevers for gas sensing.

Considering the various disadvantages mentioned in the preceding sections, most of the modern day sensor technologies are not feasible to be used for IoT based applications. The commercially available gas sensors used for such purposes are used mostly for air-quality monitoring in manufacturing, agriculture and health industries and include devices which can detect analytes such as $CO_2$, CO, $H_2$, $O_3$ and $O_2$ [16]. These are usually electrochemical, photoionization and semiconductor based sensors [17, 18]. However, many of them require complicated circuits with multiple operational amplifiers for optimum performance. Correct sensor output is limited to operational temperatures < 40° C and the devices are cross-sensitive to a multiple of commonly found gases.

There is considerable research going on in the field of low power gas sensing. Laubhan [19] proposed a low power IoT framework where they used multiple sensor chips as part of a wireless sensor network with configurable nodes which can be used to analyze motion detection, perform air quality inspection, measure humidity and temperature. Gogoi et al. [20, 21], [66], also developed batch-fabricated multisensor platforms on a single chip which are probable candidates to be used for IoT based applications. Novel sensing mechanisms implemented by Chikkadi [22], Choi [23], Park [24], Woo [25] and Shim [26], have gone a long way at developing sensors for low-power applications. However, to more effectively develop sensor node systems for such applications, active research is being carried out to develop gas sensors which consume even lesser power and demonstrate higher selectivity/sensitivity. One approach is to use quantum tunneling nanogap junctions for gas sensing purposes.

The following sections will briefly discuss the phenomenon of quantum tunneling, a brief history of its discovery and its applications in modern technology and provide a brief introduction to quantum tunneling sensors.

### 1.3 <u>Quantum Mechanical Tunneling</u>

Quantum mechanical tunneling is the phenomenon defined under the realm of quantum mechanics where a particle can pass or "tunnel" through a barrier potential greater than its energy. This is in direct contradiction with classical mechanics which states that any physical body cannot surmount a potential barrier greater than its own energy. Figure 1.6 shows the schematic representation of an electron wave tunneling through a classically forbidden energy barrier.

The concept of quantum tunneling was a result of studies on radioactivity and one of its first occurrences was reported by Robert Francis Earhart, who in 1921 noticed an unexpected conduction regime when he was trying to understand conduction of gases between closely placed electrodes [27]. In 1926, Franz Rother, used a sensitive platform galvanometer to study field emission currents between closely spaced electrodes in high vacuum [28]. But it was in a paper published in 1927 by Friedrich Hund named "*Zur Deutung der Molekelspektren. I*", roughly translated means "To the interpretation of the molecular spectra", where he discussed the phenomenon of quantum tunneling to explain the outer electron moving in atomic potential with two or more minima in potential energy separated by a classically impenetrable potential barrier [29]. The next major milestone in the history of quantum tunneling was achieved by Lothar Nordheim with the help of Ralph Fowler where he calculated the transmission probability of an electron wavefunction across a steep potential and showed that either reflection or transmission of the wave could occur with nonzero probabilities, whereas classically either one of the two can occur [30]. This is famously known as the Nordheim Fowler regime of electron tunneling where an electron wave tunnels through a triangular barrier under a high bias voltage. Later in 1927, Oppenheimer and also in March 1928, Fowler and Nordheim provided the analysis of transmission rate of an electron across a triangular barrier and proved the exponential dependence of tunneling probability on both barrier width and height. It was in the year 1930 that quantum tunneling proved most critical, when Oscar Rice interpreted the tunneling phenomenon as an analogy to alpha decay [31]. Rutherford, who was till then confused with the results he got from his scattering experiments (he observed that the α-particles which were emitted from U-238 were known to possess an energy of 4.2 MeV whereas the Coulomb potential was greater than 8.57 MeV), would eventually thank George Gamow, Ronald Gurney and Edward Condon who explained the phenomenon of alpha particle decay using quantum tunneling. About three decades later, Leo Esaki, Yuriko Kurose and Takashi Suzuki invented the tunnel diode (also known as the Esaki diode) which exploited the quantum tunneling phenomenon and displayed negative differential resistance (NDR) [32]. These were used as oscillators and high-frequency trigger devices and were noted for their extreme longevity. Devices fabricated in the 1950s are still functional. A few years later in 1963, John G. Simmons provided an analytical model describing tunneling current between two metal electrodes separated by a thin dielectric film [33]. This mathematical model has been widely used to describe tunneling current in solid-state devices and it also lies at the heart of our devices. The modern day world has much to thank Hund for his efforts in discovering the mystical phenomenon of quantum tunneling. Scanning Tunneling Microscopy (STM), Attenuated Total Reflection (ATR) spectroscopy method, Flash-memory are just some of the precious gifts of quantum tunneling.

## 1.4 Tunneling Transducers

Tunneling across a vacuum barrier was investigated by Binnig and Rohrer as part of their work which lead to the invention of the Scanning Tunneling Microscope [34]. The device consisted a piezo element which had a metal-tip fixed at the end. The tip was brought very close (in the z-direction) to the surface under observation which resulted in electrons tunneling across vacuum and a measurable tunneling current. A control-unit applied a voltage to the piezo element to maintain the tunneling current while the tip scans the surface in the x and y direction. Therefore, assuming the barrier height to remain constant, the applied voltage which was required to maintain the tunneling current was a function of the topography.

In the early 1990s, as part of his work with the Nasa Jet Propulsion Laboratory (JPL), Dr. William Kaiser developed a new class of micromachined sensor technology based on the quantum tunneling. Taking inspiration from the STM, he wanted to exploit the exponential dependence of tunneling current on potential barrier height and tunneling distance. This would allow these sensors to display much higher sensitivities than their capacitive or piezoresistive counterparts. As part of this work, Dr. Kaiser built the world's first micromachined tunneling accelerometers and tunneling IR sensors [35-38] as shown in Figure 1.7. Essentially, the accelerometer device consists of a tip-ended micromachined cantilever suspended on top of strategically placed electrodes. The cantilever is electrostatically pulled down by applying an appropriate bias voltage on one of the bottom electrodes. When the distance between the tip and the other lower electrode is near ~1 nm, tunneling current can be measured and detected by a feedback circuit. This feedback system controls the deflection voltage to maintain the position of the suspended cantilever. When the device experiences acceleration, the bias voltage (controlled by the feedback loop) maintains the relative position between the tip and the lower electrode. This bias voltage is recorded as output signal for the accelerometer.

These devices consist of a small chamber filled with gas at atmospheric pressure. The chamber lies between a pair of thin $Si_3N_4$ membranes and consists of a suspended IR absorber membrane in the middle. Before exposure to IR radiation, a tunneling bias is applied to bring the deflection electrodes towards the tip. At a tip distance of ~1 nm, a small tunneling current of ~1.5 nA is measured. When IR radiation is incident on this device, it enters through the thin top membrane and gets absorbed by the membrane. This heats the air which expands and tries to lower the membrane to deflect closer to the tip, as shown in the schematic. A feedback system ensures that the tunneling current remains 1.5 nA by appropriately controlling the deflection voltage. This deflection voltage is the transducer output and is a measure of the IR radiation. Similar to such devices, Richard Colton built a tunneling magnetometer [39]. The operation of the device was similar to those described above. In this device, a magnetostrictive ribbon (Metglas 2605SC) was used to detect small changes in magnetic fields. The tunneling current between the ribbon and the metallic tip was maintained at a constant value by applying an appropriate displacement voltage. This voltage was the output signal for the magnetometer.

The devices described in this section are some of the first and premier efforts in utilizing quantum tunneling for highly sensitive transduction. However, they were mainly physical sensors which were used as motion-sensors or IR detectors. These devices required an initial high bias voltage (greater than 120 V) to reduce the air-gap between the tunneling junctions. In addition to this, sensor operation also required a sophisticated feedback system to ensure proper working of the device. The working principle and device design also suggest that they might be susceptible to temperature fluctuations.

To overcome some of these challenges, one can use nanogap electrodes for building tunneling junctions.

## 1.5 Nanogap Electrodes

Nanogap electrodes can be defined as a pair of electrodes, separated by a gap of just a few nanometers. One of the first methods to realize a nanogap was the Mechanically Controllable Break junction (MCB) which was used by M.A Reed when he was investigating the conductive properties of di-thiol molecules [40]. In an MCB junction, a notched metallic wire is glued to an elastic substrate. The substrate is bent with a piezoelectric element which causes the wire to fracture. After this, the distance between the two wire segments can be brought closer together by the piezo element to form a nanogap between the wires. Electrochemical methods are a simple way to fabricate nanogap electrodes. In this method, nanogaps are reversibly formed by controlled chemical deposition of specific atoms on lithographically defined nanogap electrodes to close down the gap between them. Dolan [41] established the oblique angle shadow evaporation method where an elevated mask in combination with an angled metal deposition is used to define metal leads with nanometer spacing. Electromigration, which has been infamously identified as a failure mode in the electronic industry, has been successfully used to fabricate nanogap electrodes [42]. In this method, a large current is passed through an e-beam lithographically (EBL) defined metal nanowire, which leads to electromigration of the metal atoms and eventual breakdown and fracture of the thin wire, leading to formation of a nanogap. Hatzor [43] introduced a new method where they used mercaptoalkanoic acids to pattern nanowires. Here, subsequent coatings of metal-organic resist on top of EBL metal patterns lead to a controlled gap between neighboring mercaptoalkanoic layers. Metal evaporation into the gap and the subsequent lift-off of the resist layer leads to a well-defined metal pattern, thereby forming a nanogap.

While nanogap electrodes fabricated by each of these methods have resulted in valuable results, which have gone a long way into understanding essentials of charge transport across break junctions, there are also certain fundamental disadvantages in using them [44]. For example, the MCB method is too cumbersome for high-density circuit applications since it requires macroscopic piezoelectric components for nanogap formation. Electrochemical methods require precise feedback mechanisms in real-time to monitor and accurately fabricate the electrodes with a precise nanogap between them. The oblique angle shadow method requires very low-temperature conditions for metal evaporation resulting in small metal grain sizes (thereby ensuring a uniform control of the nanogap between the electrodes). Electromigration essentially requires joule heating to form the nanogap, which means that there is also a high chance of undesired melting of metal. Also, electromigration sometimes leads to deposition of debris at the critical junction. Other methods involve expensive fabrication techniques like EBL and Molecular Beam Epitaxy (MBE) for consistent results. Furthermore, most of the nanogaps formed using these methods are nonuniform in nature, which makes them unsuitable for bio-sensing applications.

To realize robust, uniform, tunable and CMOS compatible molecular tunnel junctions which can cater to detection of a variety of organic molecules, sidewall etched nanogap tunneling electrodes were introduced in 2006 [45-47]. Essentially, these devices consist of a top and bottom pair of electrodes electrically isolated by a thin insulating dielectric spacer layer. The spacer is partially etched away along the edges, wherein after chemical functionalization, organic molecules end up covalently attached to the electrode pair. These newly attached molecules provide additional electrical pathways for charge conduction between the electrodes. Therefore, this allows for inspection into charge transport across the molecular junction with and without conduction paths introduced by the foreign molecules as illustrated in the schematic of Figure 1.8, effectively decoupling the electrical characteristics of the covalently bonded molecules and the platform device. The functional molecules (for example a SAM of thiols) can be localized to desired locations between the nanogap electrodes to form the metal-molecule-metal junctions.

We introduce a new batch-fabrication method of nanogap tunneling junctions and perform exhaustive characterization of the uniformity of the spacer layer, inspect tunneling current characteristics and determine the potential barrier of the thin spacer layer as well as maximum operating voltage for the device. Molecular devices based on this construction method have been previously used as tunneling chemiresistors for bio sensing [48-51] and are potential candidates for low-power consumption gas sensing devices [52-55] [67-73]. The nanogap electrodes are chemically functionalized by coating them with a self-assemble-monolayer (SAM) of thiol molecules. When the functionalized devices are exposed to the target molecules, they get "captured" by the SAM. The captured molecules for a molecular bridge across the junction producing an augmented electrical transport between the electrodes. Therefore a nanogap structure can be utilized for electrical detection of bridging target molecules

Nanogap electrodes have been previously used to investigate the phenomenon of charge transport and bio-sensing [48, 49, 56-65]. They are potential candidates for low-power, portable, selective and highly sensitive sensor applications. Here, we present a new family of chemiresistive gas sensors that are based on electron quantum mechanical tunneling between two metal electrodes separated by nanogap junctions. In this device, shown in the schematic of Figure 1.9, two gold electrodes separated by a ~6 nm-thick gap are coated with a self-assembled monolayer (SAM) of conjugated thiols. The resulting SAM-coated gap is designed such that if no gas is captured the tunneling resistance is high, in the order of $10^9\ \Omega$ corresponding to the device's "OFF" state. If however the SAM captures target gas molecules, the resulting metal-SAM-molecule-SAM-metal junction forms a molecular bridge between the otherwise electrically isolated electrodes. Figure 1.9 shows a schematic of the sensor action focusing on the device structure and target capture corresponding to an electrical switch which is "OFF" and then turns "ON," respectively.

Using a similar fabrication method, we also built nanogap devices and used them for label-free conductivity based detection of two different types of proteins – Carbonic Anhydrase (CA-2) and Bovine Serum Albumin (BSA). The junction resistance, which is in the order of $10^9$ $\Omega$ after device fabrication, reduces to lower than 10 $\Omega$ after successful capture of the protein bio-molecules.

We also present the working of a new type of microfabricated quantum tunneling hygrometer that is able to provide large output range and a low temperature dependence. Figure 1.10 shows the schematic explaining the working principle of the humidity sensor. The device utilizes the expansion of a polymer that swells when exposed to humidity as in a polymer swelling based capacitive humidity sensor, but it produces a resistive output that measures the polymer expansion through tunneling current across a humidity dependent, thermally stabilized nanogap. The tunneling current changes many orders of magnitude providing similar output as the resistive type device with a low temperature dependence.

## 1.6 <u>Dissertation Outline</u>

The main focus of this dissertation is to develop a new generation of ultra-low power and highly sensitive MEMS sensors/switches. These devices are also engineered to display augmented selectivity in comparison to other state-of-the-art sensing techniques. The subsequent chapters explore device characteristics and sensor response to the target analyte.

Chapter 2 discusses the underlying concept of quantum tunneling which lies at the heart of these devices. A brief history of its discovery, interpretation and immense impact on how we view the physical world is also presented in the chapter.

Chapter 3 introduces the reader to micromachined quantum tunneling sensors. It presents a brief literature survey of previously fabricated devices and their device operation. It also talks about the various advantages and disadvantages of using such devices.

Chapter 4 provides an extensive statistical analysis of ultra-thin films (<10 nm to determine its applicability to batch-fabricate these devices.

Chapter 5 describes the design, fabrication and electrical characterization of the batch-fabricated devices.

In Chapter 6 we present the device response to target analyte. We also include device response to other analytes, thereby commenting on the selectivity of the gas sensor.

Finally, in Chapter 7, we try to describe the device response using preexisting analytical models of quantum tunneling transport. The analytical equation is used to investigate the mode of current conduction in the device and help provide a thorough explanation of its working.

## 1.7 <u>References</u>

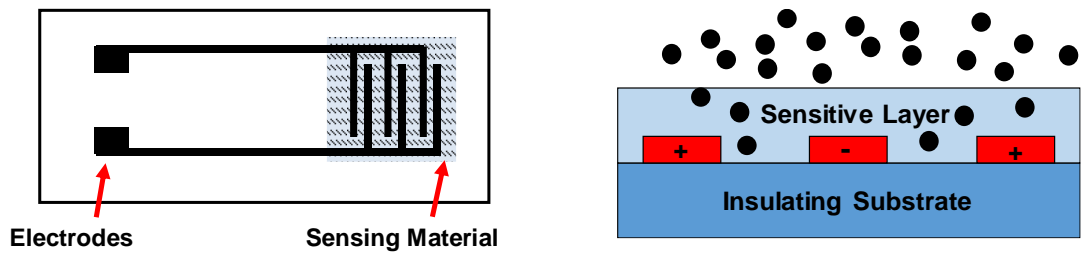

**Figure 1.1**: Schematic showing the layout of a conductivity sensors.

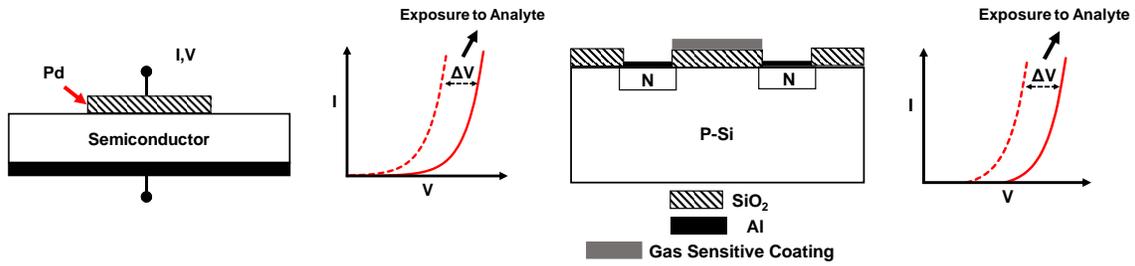

**Figure 1.2**: Schematic showing the working principle of solid-state gas sensors.

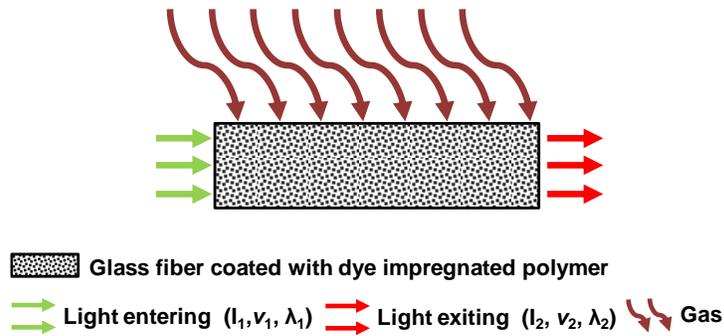

**Figure 1.3**: Schematic of an optical gas sensor showing its working principle.

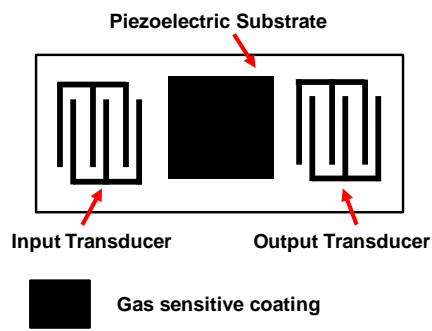

**Figure 1.4**: Schematic representation of piezoelectric gas sensors and electrode layout.

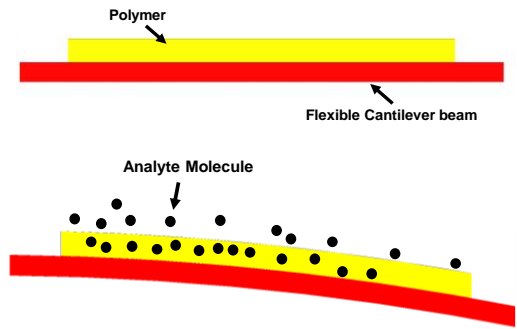

**Figure 1.5**: Schematic showing the bending of MEMS cantilevers upon exposure to analyte.

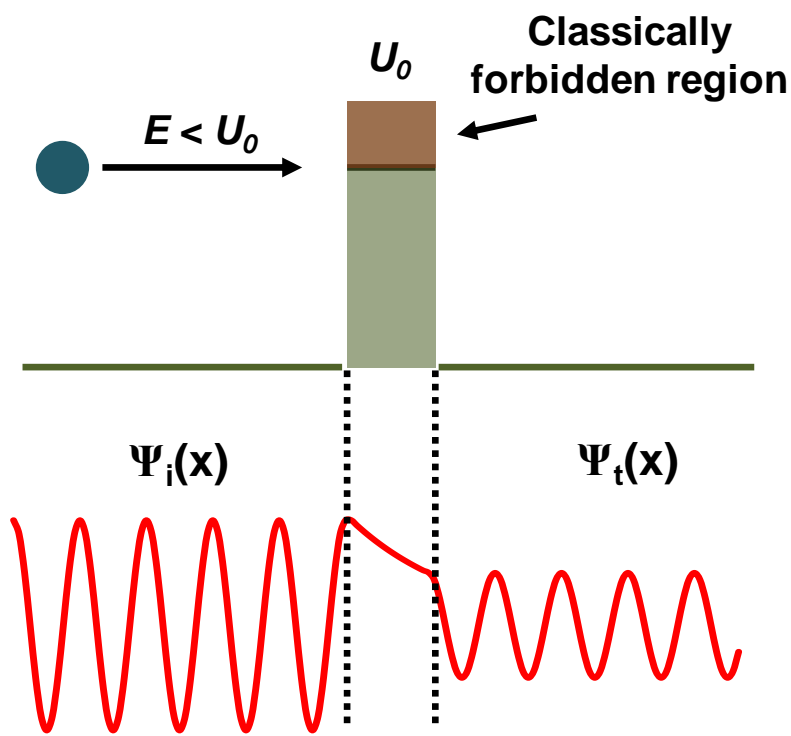

**Figure 1.6**: Representative diagram showing quantum tunneling of particle wave across a classically forbidden potential barrier.

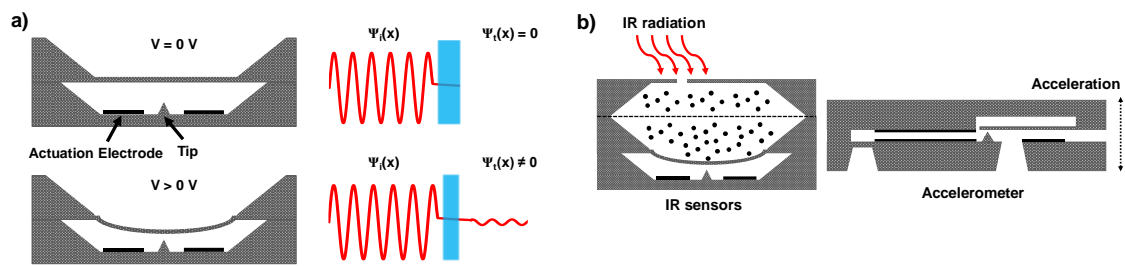

**Figure 1.7**: Preliminary tunneling sensors built by Dr. Kaiser. a) Working principle of the tunneling transducers. b) Schematic depicting the tunnel sensors .

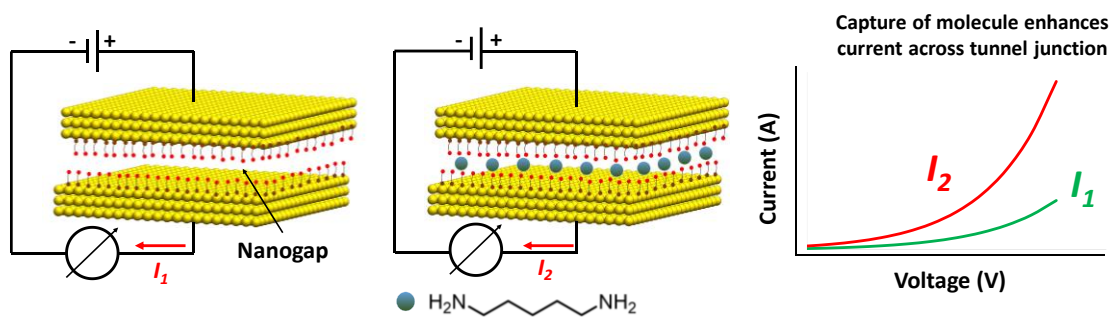

**Figure 1.8**: Schematic of tunneling nanogap junctions. Trapping the molecules within the nanogap alters the junction I-V characteristics.

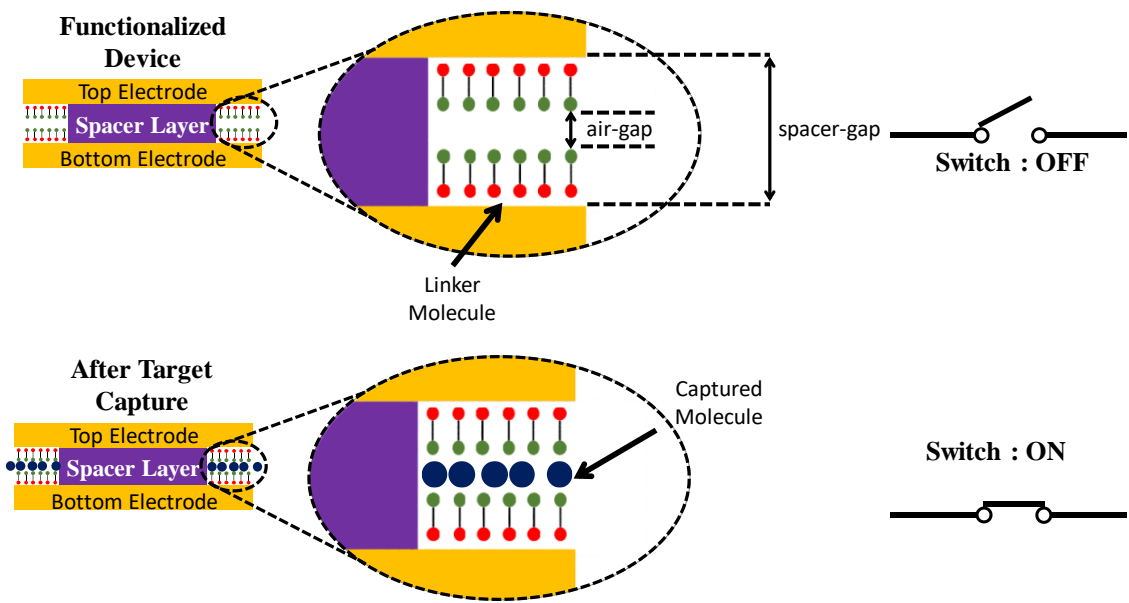

**Figure 1.9**: Schematic of nanogap device after fabrication and analyte capture. Successful target capture turns the switch ON.

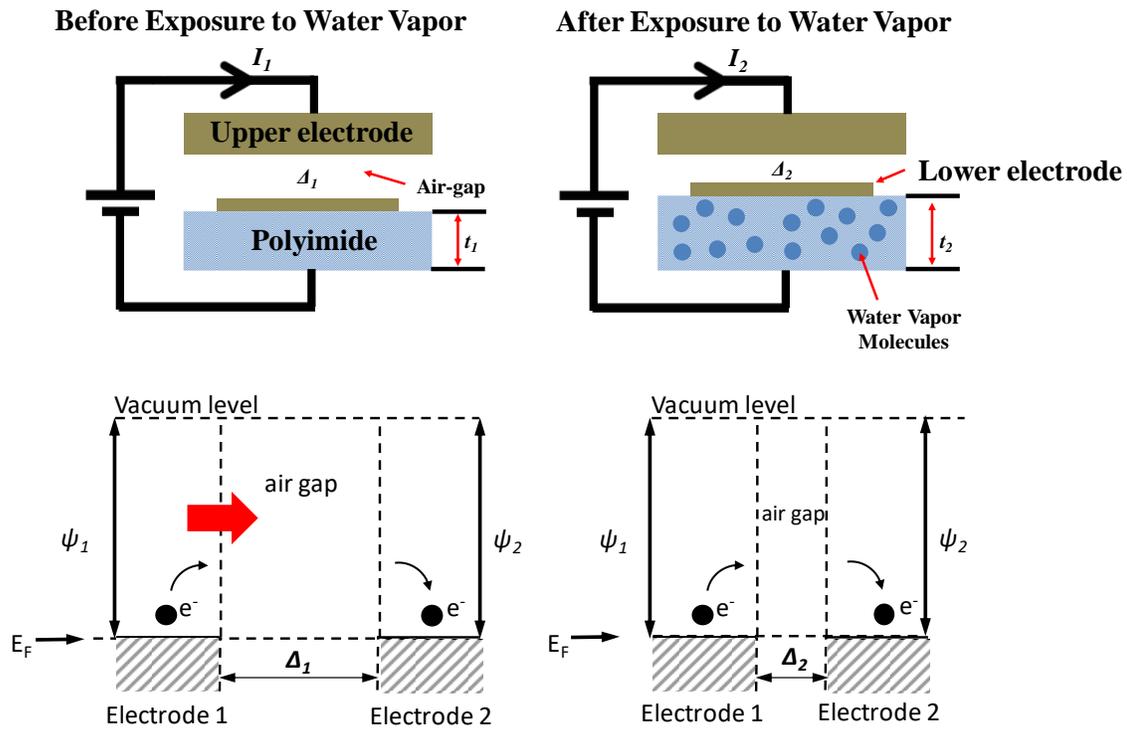

**Figure 1.10**: Working principle of the tunneling humidity sensor. Exposure to water vapor molecules reduces tunneling distance.

# BATCH-FABRICATED α-Si ASSISTED NANOGAP
# TUNNELING JUNCTIONS

## 2.1 Introduction

In this chapter, we describe a new method to fabricate vertical nanogap tunneling junctions where a spacer layer is used to define the thickness of the break junction. We also report statistical data concerning thickness uniformity of the spacer films, quantum tunneling current characteristics of the fabricated devices and experimentally determine the potential barrier of the spacer layer using transition voltage spectroscopy (TVS).

## 2.2 Experimental Procedures

### 2.2.1 Thin-film deposition and characterization

The uniformity and repeatable deposition of the spacer stack are crucial to a high-yield batch-fabrication of the nanogap electrodes. To determine the deposition rate and perform rigorous statistical analysis of the uniformity of these ultra-thin films, we deposited six different thicknesses (3–8 nm) of $SiO_2$ on 4-inch Si wafers as described in 2.2.3 below. Every variant of $SiO_2$ thickness was deposited on five wafers

---

each. We then performed thickness measurements on sixty-nine sites on each of these thirty samples. Similarly, different thicknesses of α-Si thin-film were deposited on 4-inch Si wafers. Due to ellipsometry modeling restraints, a layer of 1 μm of thermal $SiO_2$ had to be grown on the sample before the α-Si was deposited. Standard optical methods were used to determine the thickness of both the thin films on the n&k Analyzer 1500 D (n&k Technology). AFM measurements were performed on lithographically patterned thin films to experimentally determine their surface roughness as well as that of the substrate.

### 2.2.2 Device structure, design and fabrication

The nanogap electrode assembly consists of a partially etched spacer film, sandwiched between two thin electrically isolated gold electrodes. The spacer is a sacrificial stack of a very thin dielectric layer of $SiO_2$ deposited using a plasma-enhanced ALD method, which provides excellent electrical insulation and an ultra-thin layer of sputtered α-Si, which acts as an adhesive layer between the top gold electrode and the dielectric material. A sacrificial plasma etch of the spacer layer creates a nanogap along the edges of the upper gold electrode. A schematic of the fabricated two-layer nanogap design is shown in Figure 2.1. Since these nanogap devices will eventually be used for low-power and remote gas-sensing, it is essential that the leakage current during device operation be kept to a minimum so that the parasitic DC power-consumption is extremely low. Since the junction leakage current is directly proportional to the overlap area of the electrodes, an electrode design with a low overlap area should generally ensure a lower leakage current. Therefore, we chose two design architectures having relatively low electrode overlap areas—a "square-overlap" layout (having an overlap area of ~16 μm²), which is essentially a perpendicular arrangement of two thin metal wires and a "point-overlap" layout (having an overlap area of ~0.24 μm²), which is a very low-overlap arrangement of lithographically patterned pointed tip-ends of patterned electrodes. This also allowed us to investigate current conduction characteristics as a function of the overlap area.

Figure 2.2 shows the fabrication process of the nanogap electrode assembly. We start by growing ~300 nm of $SiO_2$ on a Si wafer (a). This is followed by DC sputtering 25 nm of Cr and 200 nm of Au and subsequent patterning by traditional lithographic techniques to define the lower gold electrodes (b–c). The chemical solution Transene Au etchant TFA was used to selectively etch away the gold. Next, a desired thickness of dielectric material ($SiO_2$) was deposited for various time intervals, from 17 to 188 cycles of plasma-enhanced ALD process at a substrate temperature of 200 °C with the commercially available metal-organic precursors tris[dimethylamino]silane (3DMAS) on separate samples to fabricate nanogap electrodes with different spacer thicknesses (e). Then, an ultra-thin layer of α-Si was sputtered for 17 seconds at 50 W to get a ~1.5 nm film on each sample. Without breaking vacuum, another layer of ~200 nm of Au was sputtered and lithographically patterned to form the top electrodes (f–g). Finally, the samples were dry etched in an inductively coupled plasma etcher (Oxford 100 ICP) with $SF_6$ plasma for 40 seconds at an ICP forward power of 250 W with 45 sccm of $SF_6$ flow rate to partially remove the α-Si and $SiO_2$, thereby forming a nanogap along the edges of the top electrode (h).

### 2.2.3 Choice of spacer stack: $SiO_2$ as the dielectric material and α-Si as an adhesive layer

Since the intended application of the fabricated nanogap sensors is chemical detection and resistance switching at very low standby DC power, the primary requirements of the chosen dielectric material are extremely low leakage current, very high off-resistance and compatibility with standard CMOS fabrication techniques. Keeping these factors in mind, plasma enhanced ALD $SiO_2$ was chosen as a dielectric material since for a given geometry and thickness of dielectric film, and operating bias voltage, the leakage current is lowest for $SiO_2$ films [1].

Since Au is a noble metal, it is chemically inert and does not easily form oxides. Therefore, Au does not adhere well to dielectric films like $SiO_2$, which is widely used in CMOS processes. Although sputtering of Au at elevated temperatures on $SiO_2$ substrates is an effective solution [2], the most common practice is to deposit a thin metallic adhesive layer of Cr, Ti or Ni before depositing the layer of Au. However, such adhesive layers have been known to cause thermal degradation of the film because of grain-boundary diffusion [2] and to ensure a very high off-resistance of the device, using a nonmetallic adhesive layer could be a possible solution. Taking these into consideration, we report a novel application of an ultra-thin layer of sputtered α-Si as an effective adhesive layer for Au in microfabrication processes.

### 2.2.4 Imaging and electrical characterization of nanogap electrodes

After fabrication of the devices, high-resolution SEM imaging was done at an accelerating voltage of 15.0 kV by the FEI Nova NanoSEM to inspect the gap formed between the Au electrodes. I-V characteristics of the nanogap devices were measured on the Keithley 4200A-SCS Semiconductor Parameter Analyzer to ensure that the electrodes were electrically isolated after the nanogap formation and that the resistance between them was substantially high before chemical functionalization and exposure to target analyte. To ensure low-noise and high fidelity electrical signals, tunneling current measurements were performed in a dark room, inside an electrically shielded enclosure using the Keithley Parameter Analyzer. Instantaneous breakdown voltages of each of these devices were experimentally determined using a simple voltage ramp-up test where the biasing voltage across the Au electrodes was increased at a constant ramp-up rate until the dielectric stack suddenly began to conduct electricity.

## 2.3 Results and Discussion

### 2.3.1 Sacrificial film characterization

Figure 2.3a and 2.3b show an approximately linear deposition rate of ~0.7 Å of $SiO_2$ per cycle of plasma-enhanced ALD at 200 °C and a deposition rate of 0.9 Å per second for sputtered α-Si. Figure 2.3c-d show AFM images of surface topography of the patterned features of α-Si and ALD $SiO_2$ thin films over scan area of 100 μm². The average surface roughness of the α-Si, ALD $SiO_2$ and the substrate was measured to be ~35 pm, ~34 pm and ~20 pm.

Figure 2.4a shows the contour mapping of $SiO_2$ film thickness on a 4-inch wafer for six different deposition cycles and standard deviation measurements of each of its five repetitive depositions are given in the Figure 2.4b. The interpolated contour plots and standard deviation data were obtained from using the JMP statistical analysis software developed by the SAS Institute. The maximum standard deviation, which is a direct measurement of the film's nonuniformity, was found to be ~6 Å. Figure 2.5 shows the variation in film-thickness for each repetition. The maximum standard deviation in thickness across multiple depositions was 5.57 Å (for 48 cycles of ALD). Similar experiments were performed to characterize the α-Si film, which was also deposited on a 4 inch oxidized Si wafer. Measurements indicate a standard deviation of 5.73 Å on the sample. These measurements are a clear indication that there is minimal variation in spacer layer thickness. Therefore this fabrication technique can be effectively used for batch fabrication of the nanogap electrodes with sub 10 nm spacer thickness.

### 2.3.2. SEM imaging

SEM images of the fabricated square-overlap layout and point-overlap layout devices are shown in Figure 2.6(a –b), respectively. The device footprint for the square-overlap and the point-overlap schemes are ~0.36 mm2 and 0.27 mm2 and a zoomed in section of their overlap region reveals an area of ~16 µm2 and 0.24 µm2 respectively. SEM images of the nanogap between electrodes for various spacer thicknesses are shown in Figure 2.7(a-c). As evident by the SEM images, the nanogap thicknesses are in good agreement with the optical measurements discussed in 2.3.1.

### 2.3.3. I-V characteristics of square-overlap and point-overlap layout devices

Figure 2.8 (a-b) and (c-d) show I-V characteristics of square-overlap and point-overlap device respectively for various spacer thicknesses. As evident from the plots, the current exponentially reduces as the thickness of the spacer layer increases from 4 nm – 6 nm. Also, the junction current is significantly lower for the point-overlap in comparison to the square-overlap layout due to a significant reduction in overlap area. The average junction resistance of the square-overlap layout devices ranged from 3.22 GΩ (for 4 nm spacer layer thickness), $3.83\text{x}10^3$ GΩ (for 5 nm spacer layer thickness) and $33.3\text{x}10^3$ GΩ (for 6 nm spacer layer thickness) and the average resistance of the point-overlap layout devices ranged from 25 GΩ (for 4 nm spacer layer thickness), $1\text{x}10^4$ GΩ (for 5 nm spacer layer thickness) and $46\text{x}10^3$ GΩ (for 6 nm spacer layer thickness).

Figure 2.9 (a) shows the I-V characteristics on a square-overlap layout design having spacer thickness of 4 nm for 20 repetitive cycles. As evident from the plot, there is negligible change in the I-V characteristics even after 20 repetitions of the I-V measurements. Figure 2.9 (b-d) show I-V characteristics of the nanogap electrode devices across the wafer. The plots indicate a maximum variation of one order of magnitude in the I-V curves. The reason for that is an exponential dependence of tunneling current on spacer gap. As shown in Figures 2.4 and 2.5, the standard deviation of the deposited spacer films is ~0.5 nm. Therefore, the tunneling current is susceptible to a maximum variation of 2 order of magnitude across the wafer.

The plot shows that there is no significant change in I-V characteristics over different regions of the wafer. The differences in I-V curves are a result of minor nonuniformities in thickness and surface defects of ultra-thin films deposited over 4-inch wafers. However, as the plots indicate, this method can be effectively used to batch-fabricate nanogap electrodes having a gap of <10 nm.

### 2.3.4 Tunneling current measurements and transition from direct tunneling to Fowler-Nordheim tunneling

Different charge transport phenomenon can occur across a junction barrier as function of barrier height and thickness, operating temperature and biasing voltage. At low junction widths and very low biasing voltages, if the temperature is sufficiently high, classical charge transport can occur. Charge carriers can overcome the barrier potential because of thermal energy at high ambient temperatures. This regime of conduction is the "thermionic emission" [3]. When the ambient temperature is low, there cannot be any classical charge transport across the barrier. However, with an increasing biasing voltage, the barrier potential becomes approximately trapezoidal and "direct quantum tunneling" is the dominant charge transport phenomenon across the barrier potential [3]. When the bias voltage is increased further, the barrier potential becomes approximately triangular and the charge transport phenomenon is described by Fowler – Nordheim (F-N) tunneling [3]. The differences in barrier shape between Direct Tunneling and Fowler Nordheim tunneling regimes are shown in Figure 2.10 (a-b).

Tunneling I-V characteristics of a Metal-Insulating-Metal (M-I-M) device can be expressed by the following equations [4, 5] : $I = V \times exp^{-(\frac{2d\sqrt{2m\Phi}}{\hbar})}$ (for V < $V_{trans}$ : direct tunneling) and $I = V^2 \times exp^{-(\frac{4d\sqrt{2m\Phi^3}}{3heV})}$ (for V>$V_{trans}$ : Fowler-Nordheim tunneling ), where d is the insulator layer thickness, e is the charge of an electron, and $V_{trans}$ is the voltage at which the tunneling current regime changes from direct-tunneling to Fowler-Nordheim tunneling, and it is approximately $V_{trans} = (2/3) \times (\Phi/e)$. For the same current level, at larger thicknesses the F-N current becomes dominant because V>$V_{trans}$.

Current conduction across the composite stack layer of partially etched away α-Si and SiO$_2$ can be modelled using the Simmon's approach for any arbitrary barrier shape [6]. It has been demonstrated that Joshi [7] and Ikuno [4] that the potential barrier height of a dielectric film can be deduced by plotting the transition of tunneling current regime from Direct Tunneling to Fowler Nordheim current. Figure 2.10 (c) shows the transition from Direct tunneling to Fowler Nordheim tunneling at ~0.28 (V$^{-1}$) or ~3.5 V, which is the transition voltage for devices having a spacer thickness of 6 nm. Therefore, the spacer stack of the device has a potential barrier of ~3.5 eV. This value is similar to the results obtained by Joshi [7] for SiO$_2$ films. The linear slope of the graph beyond the threshold voltage and logarithmic slope at voltages lower than the threshold clearly indicate that Direct Tunneling occurs at low bias voltages and Fowler Nordheim tunneling occurs at higher biasing voltages through the composite spacer stack of the fabricated device.

### 2.3.5 Dielectric breakdown measurements and I-V measurements at various temperatures

Dielectric breakdown refers to the dielectric layer losing its insulating properties and becoming electrically conductive and is one of the major causes of device failure in the semiconductor industry. Therefore, one of the critical parameters of device characterization is the dielectric lifetime. There are mainly two failure modes observed in thin films [8]. The first is instantaneous breakdown (where the charge transport across the dielectric junction instantaneously rises very sharply when the biasing voltage reaches a critical level) and the second is time-dependent dielectric breakdown or TDDB (where the eventual breakdown of the insulating film after a specific duration of time results from a continuous charge transport across the junction). Since we are mainly concerned with determining the maximum operating voltage of the device, in this paper, we limit our discussion to instantaneous failure of the dielectric thin film. Once the breakdown voltage is applied across the device electrodes, we observe an irreversible degradation of the spacer film. Figure 2.11a shows the I–V characteristics of a fabricated ~6.0 nm spacer layer nanogap device, measured in atmospheric conditions and at room temperature. As evident from the plot, at ~ 7 V the current flowing across the device suddenly jumps to a much higher value, thereby indicating an instantaneous breakdown of the dielectric layer. Figure 2.11b shows the I–V curve of the same device after dielectric breakdown has been observed. The plots suggest that the dielectric film is now irreversibly damaged and therefore highly conductive, displaying typical ohmic behavior.

Figure 2.12a shows an almost linear dependence of the experimentally determined breakdown field value of the nanogap electrode device on spacer film thickness. The experimental data suggests that the maximum voltage which can be applied across the nanogap electrodes ranges from 2.9 V for the 4.0 nm spacer layer to 10.2 V for the 9.0 nm spacer layer, suggesting that the spacer film consisting of ∝-Si and $SiO_2$ has an average breakdown field value of ~11.0 MV/cm. Since the ∝-Si is ultra-thin, it can be assumed to be highly conductive. Therefore, it is the dielectric component of the spacer film ($SiO_2$) which degrades irreversibly. According to the thickness calibration curve of the $SiO_2$ films and the dielectric breakdown plots, the breakdown field of the dielectric layer was determined to be 13–14 MV/cm, which is in course agreement with the experimentally determined results shown by Usui et al. [9]. The minor differences in breakdown field values can be attributed to nonuniformities in deposited films.

Figure 2.12b shows the I–V traces for the square-overlap layout device having a 4 nm spacer for different heating temperatures from 30 degrees Celsius to 80 degrees Celsius. As shown in the plot, for bias voltages between –1 to +1 V, since the current conduction is mainly through tunneling, it is fairly independent of operating temperature. However, for bias voltages <1.5 V and >1.5 V, the current conduction typically resembles Schottky current emission, where the I–V characteristics are dependent on the operating temperature.

## 2.4 Conclusions

We fabricated gold nanogap tunneling electrodes with a spacer thickness as low as 4.0 nm and performed extensive characterization of the spacer layer and the device.

Optical measurements revealed an average nonuniformity of 0.46 nm in the $SiO_2$ film and ~0.58 nm in the α-Si film. Deposition rates were found to be ~0.7 Å of $SiO_2$ per cycle of plasma-enhanced ALD at 200 °C and a deposition rate of ~0.9 Å per second for sputtered α-Si. I–V characteristics showed that the fabricated devices demonstrated a maximum DC resistance of $46 \times 10^3$ GΩ between the electrodes, which is an extremely high off-resistance for switching applications. Repetitive I–V measurements on a single device showed negligible drift in electrical characteristics. Electrical measurements performed on devices across the wafer displayed some nonuniformities in electrical properties which is a direct result of minor fabrication errors and exponential dependence of tunneling current on spacer thickness. These nonuniformities are in accordance with the extensive uniformity measurements performed on the spacer layers. Therefore the fabrication method can be used to batch-fabricate nanogap electrodes having sub-10 nm spacer thickness. Tunneling current measurements demonstrated the presence of both direct tunneling and Fowler–Nordheim tunneling regimes depending on the biasing voltage. Fowler–Nordheim plots revealed that the barrier potential for the spacer layer is ~3.5 eV. Breakdown measurements showed that the average breakdown field for the fabricated devices was 11 MV/cm. I–V measurements at different heating temperatures also displayed electrical conduction which is typical of Schottky emission. Preliminary results have already been shown where nanogap sensors fabricated using this technique have been used as chemiresistors for the near reversible detection of cadaverine gas, BSA and CA-II proteins. Therefore, these devices are ideal candidates for use in sensor nodes which are part of Internet of Things (IoT) applications, continuous monitoring, and low-power sensing.

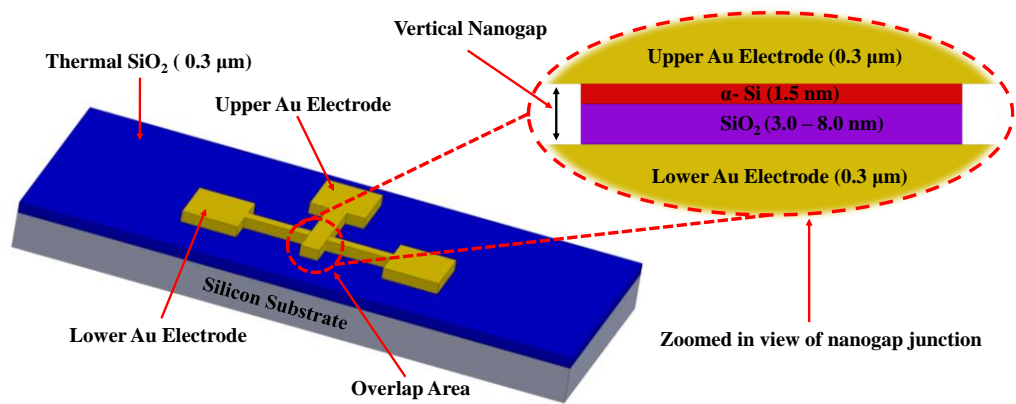

**Figure 2.1**: 3D schematic of device of vertical nanogap structure and zoomed in view of sacrificial spacer layer.

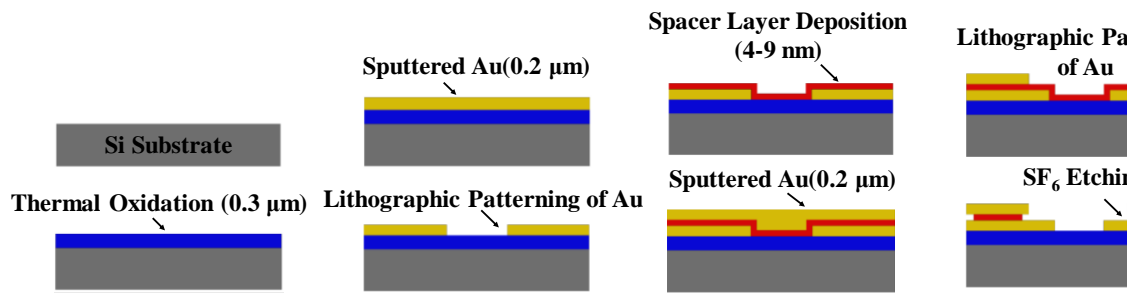

**Figure 2.2**: Simplified fabrication process of vertical nanogap electrodes separated by a thin spacer layer.

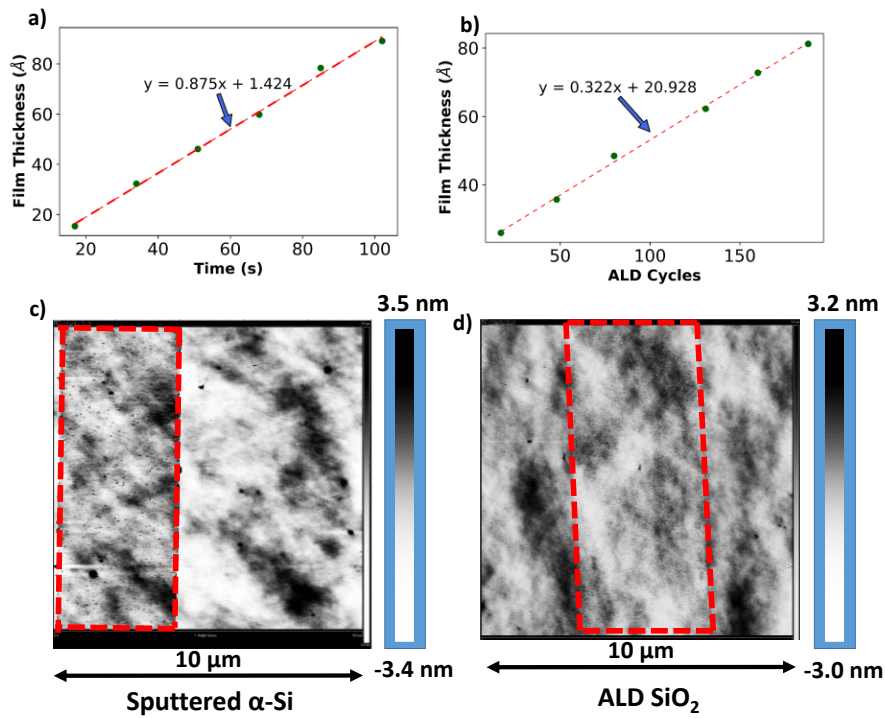

**Figure 2.3**: Thickness calibration curves for (a) sputtered α-Si and (b) ALD SiO₂ and (c) and (d) Modified AFM scans showing surface roughness of α-Si ALD SiO₂ thin films. The red dotted lines outline the features patterned using conventional lithographic techniques.

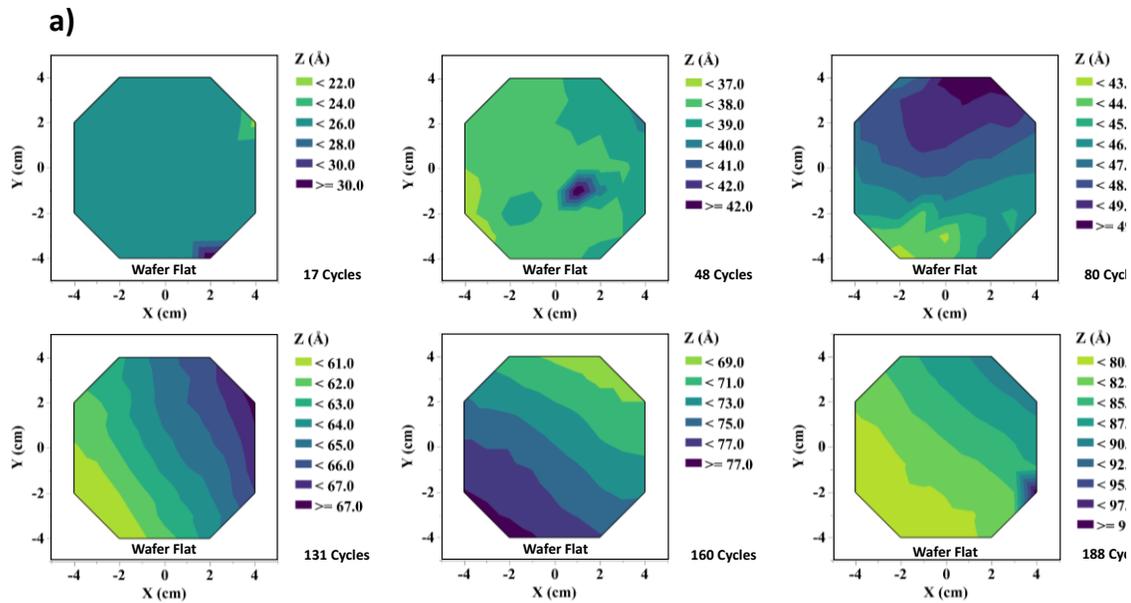

**a)**

**b)**

| No. of Cycles | Std. Dev of | Std. Dev of | Std. Dev of | Std. Dev of | Std. Dev of | Average Std. |
|---|---|---|---|---|---|---|
| (#) | Sample 1 (Å) | Sample 2 (Å) | Sample 3 (Å) | Sample 4 (Å) | Sample 5 (Å) | Dev (Å) |
| 17 | 2.51 | 1.07 | 3.59 | 2.95 | 0.98 | 2.22 |
| 48 | 0.86 | 9.49 | 0.91 | 8.59 | 9.56 | 5.88 |
| 80 | 3.9 | 1.74 | 5.95 | 4.02 | 9.32 | 5.23 |
| 131 | 5.62 | 4.63 | 1.93 | 2.77 | 7.35 | 4.46 |
| 160 | 3.23 | 6.52 | 2.70 | 3.15 | 3.23 | 3.77 |
| 188 | 8.5 | 5.61 | 3.70 | 8.18 | 4.20 | 6.04 |

**Figure 2.4**: Film thickness contour mapping, as measured by optical methods, on 4-inch wafers for different thicknesses of SiO2 and nonuniformity measurements of different SiO2 films done on five samples for each specific thickness.

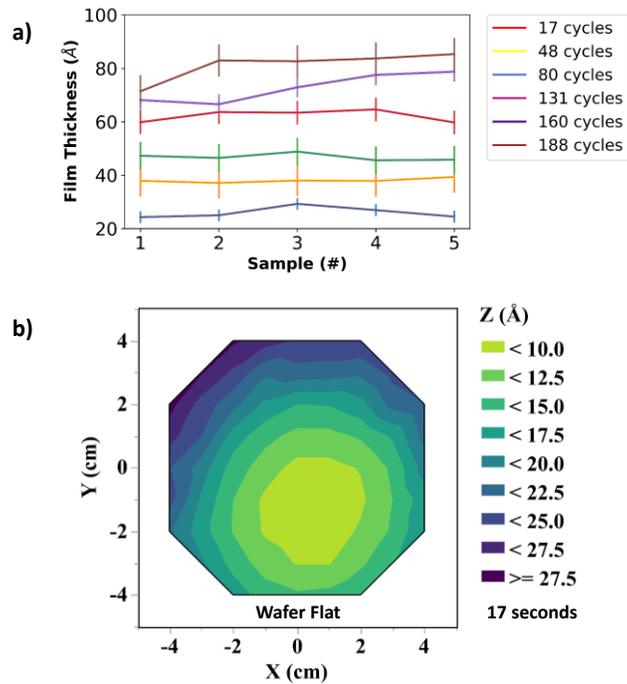

**Figure 2.5**: Uniformity measurements of (a) SiO₂ on (b) α-Si on 4-inch wafer.

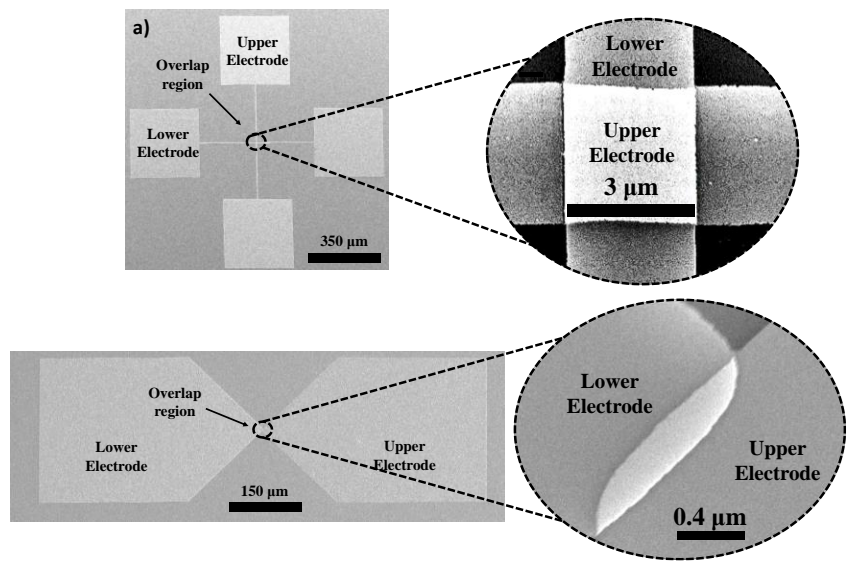

**Figure 2.6**: High resolution SEM images of (a) square-overlap layout and (b) point-overlap layout with zoomed in images of their overlap regions.

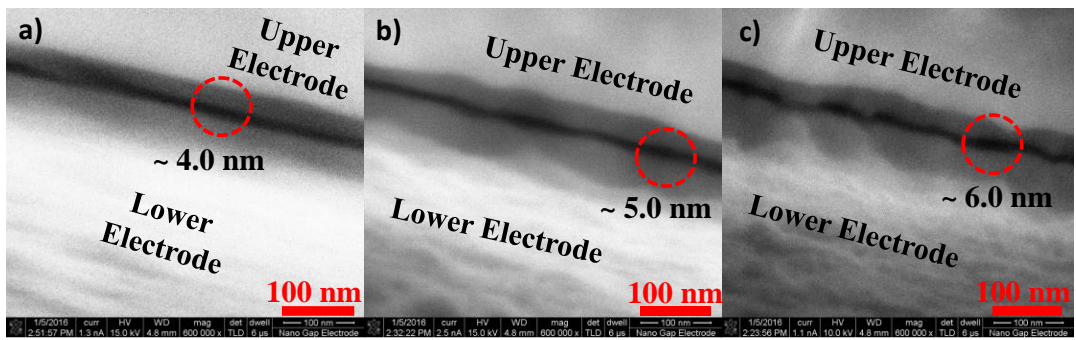

**Figure 2.7**: High resolution SEM images of (a) 4.0 nm, (b) 5.0 nm, and (c) 6.0 nm gaps.

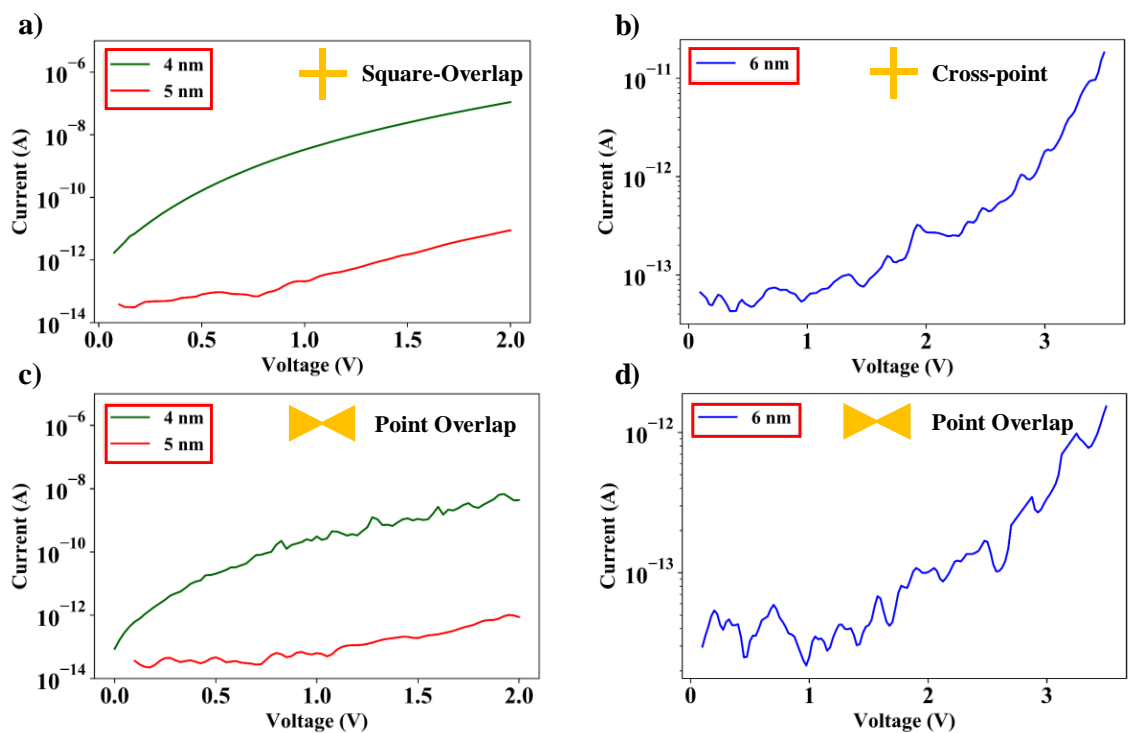

**Figure 2.8**: I-V measurements for (a) square-overlap layout having spacer thickness of 4 nm and 5 nm (b) square-overlap layout having spacer thickness of 6 nm and (c) point-overlap layout having spacer thickness of 4 nm and 5 nm (d) point-overlap layout having spacer thickness of 6 nm.

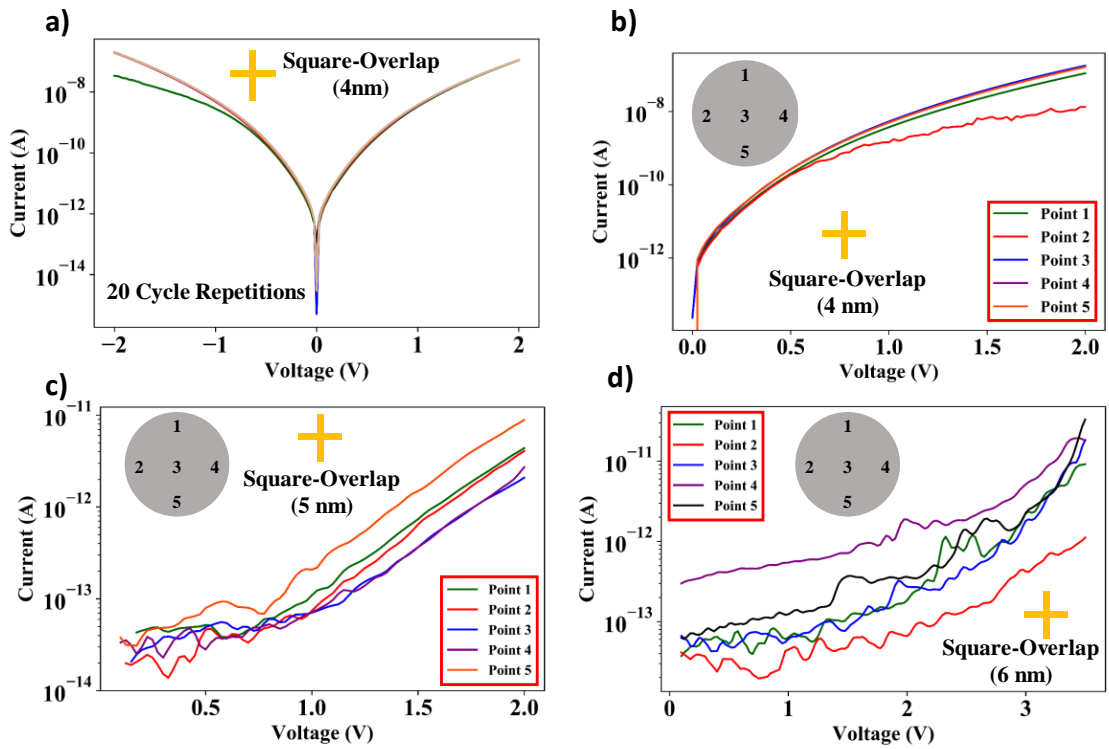

**Figure 2.9**: Plots for (a) 20 cycles of repetitive I-V measurements of square-overlap layout devices having spacer thickness of 4nm and I-V measurements of square-overlap devices over five different areas on a 4-inch Si wafer having spacer layer thicknesses of (b) 4 nm (c) 5 nm and (d) 6 nm.

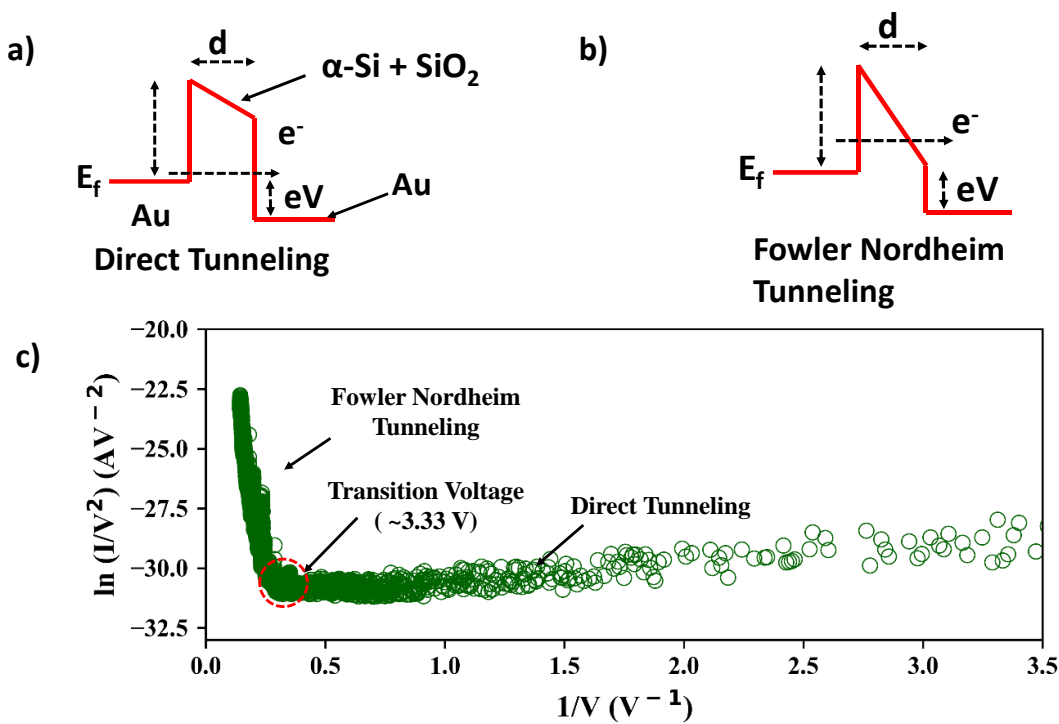

Figure 2.10: Representative potential barrier and transitional voltage spectroscopy measurements. (a) Potential barrier during direct tunneling regime and (b) Potential barrier during Fowler Nordheim Tunneling regime. (c) Transition from direct tunneling regime to Fowler Nordheim tunneling regime upon increase in biasing voltage.

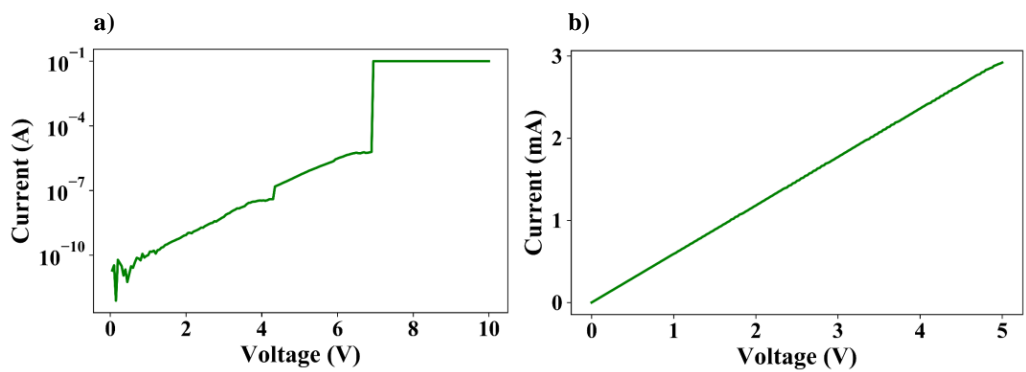

Figure 2.11: Breakdown characteristics of the fabricated nanogap electrodes. (a) Instantaneous breakdown of the dielectric film at ~7 V demonstrated by a sudden increase in conduction current. (b) I–V plot of the device after dielectric breakdown typical of ohmic conduction.

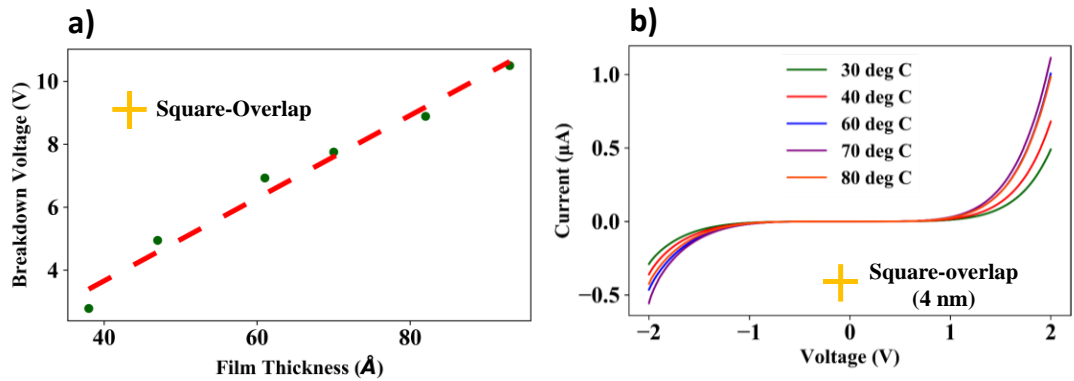

**Figure 2.12**: Device characterization and response to temperature. (a) Experimentally determined breakdown voltage vs. spacer film thickness (4nm − 9nm) for square-overlap layout and (b) I-V characteristics of 4 nm spacer thickness square-overlap layout for different temperatures.

# CHAPTER 3

# ULTRA-LOW POWER TUNNELING NANOGAP GAS SENSORS

## 3.1 <u>Introduction</u>

In this chapter, we demonstrate electrical detection of captured gases from measurements of the tunneling characteristics of gas-mediated molecular junctions formed across nanogaps in the devices fabricated in the previous chapter. We mathematically describe the sensor action using a gas-dependent tunneling device model. SAM coating procedures, coating characterization, and linker conductivity measurements have been discussed in detail. Sensor action against different concentrations of target analyte is also presented. To validate the accuracy of the presented mathematical model, we have curve-fitted the I-V characteristics of the nanogap junction after successful capture of cadaverine molecules at various analyte concentrations. Cross-sensitivity tests are carried out and presented to verify the selectivity of the nanogap sensor. We also discuss the dynamic response of the sensor and try to mathematically describe it using well-established sorption kinetic models.

## 3.2 <u>Experimental Procedures</u>

### 3.2.1 Linker solution preparation and SAM coating procedure

The SAM linker solution was prepared by dissolving about 20 mg of the linker chemical in solution of about 4 ml dimethyl sulfoxide (DMSO) and 10 ml of ethanol.

The fabricated devices were cleaned by treating them with $O_2$ RF plasma for one minute and then immediately immersed into the linker solution, to avoid any contamination. The exposed gold surface was functionalized by immersing the device in the proper linker solution for about 48 hours. This ensures perfect formation of the SAM on all exposed gold surfaces, including those sandwiching the nanogap between them. After proper functionalization, the bare gold surface was coated by a self-assembled-monolayer of (4-((4-((4-mercaptophenyl)ethynyl)phenyl)ethynyl)benzoic acid).

### 3.2.2 Sensor measurements in test chamber

The testing setup consists of a stainless steel cylindrical structure with a small opening in the top of the chamber which is sealed using a rubber septum seal. The chamber height was ~ 15 cm and the diameter of the cylindrical structure was ~5.5 cm. The opening in the top of the chamber is ~1.8 cm. The chamber is attached to an inlet and outlet so that it can be purged using Nitrogen and the chamber can be cleansed of any undesirable gases or remnant cadaverine from previous experiments. To provide electrical connection to the chamber, we used a power feedthrough electrode system manufactured by Kurt J. Lesker company and welded it to the bottom of the chamber to ensure that there would be no gas leaks. To measure the sensor activity, we wire-bonded the device to bond-pads and attached them on a glass chip platform. The bond pads were connected to electrical feedthroughs of the chamber, which were in turn connected to the probes of the Keithley 4200A-SCS using standard tri-ax cables for noise shielding. A constant DC bias of 0.7 V was applied across the electrodes and then the device was exposed to the target gas molecules. The changes in junction resistance were measured and recorded by the parameter analyzer.

### 3.2.3 Selectivity tests

We exposed our device to a few commonly found volatile organic compounds and gases such as Acetone, Ethanol, Methane and Carbon Dioxide etc. and we measured the device electrical response to each of these gases. The wire-bonded sensor was kept inside an air-tight container. To measure sensor activity, the device was connected to the Keithley Parameter Analyzer 4200A-SCS using co-ax cables and electrical feedthroughs present on the walls of the container. The analytes were then introduced to the testing chamber and resistance of our sensor was measured using standard DC-bias measurements, performed by the Keithley Parameter Analyzer.

### 3.2.4 Repeatability analysis

To investigate the repeatability of our device and monitor its recoverability, we exposed our sensor to ~ 60 ppm of cadaverine in the test-chamber. After observing the maximum possible resistance drop during the adsorption cycle, we removed the top cover of the test chamber to allow the trapped target molecules to desorb from the device and vacate the test-chamber. This was performed multiple times untill we observed a significant change in baseline resistance of the tunneling junction.

### 3.3 <u>Results and Discussion</u>

### 3.3.1 Gas-dependent tunneling device model

The electrical I-V characteristic and the device resistance is determined by the species trapped on the nanogap which can be two organic SAM layers separated by an air gap or a SAM-captured gas molecule-SAM chain. The electrical conductivity of these is examined below.

Generally speaking, the electrical properties of organic molecules can be determined by using AFM/STM [1, 2], MCB techniques [3] or can be simulated using computational chemistry methods [4-6]. Electrical transport through conventional alkane(di)thiol SAMs has been described as pure quantum tunneling across a thin dielectric film which has a rectangular potential barrier with image charge effects included, as described by J. Simmons in 1963 [7]. The dielectric constant of a typical alkane-thiol SAM layer has been determined by impedance measurements to be 2.1 according to Akkerman [8]. Akkerman also verified the barrier height of alkane-thiol SAMs sandwiched between a pair of Au electrodes and protected by a layer of PEDOT:PSS (within the spacer stack), is in the range of 4-5 eV. Although alkane-thiols are bad conductors of electricity, the conductivity of thiol molecules can also be engineered by addition of alternative alkynes and aromatic rings within the molecule itself. This modification leads to a conjugated molecular arrangement which results in an amplified charge transport along the molecule on account of delocalized π-electron orbitals, which can freely move along the whole molecule. Bower also verified that conjugation in SAMs containing oligophenyl groups leads to a reduction in attenuation constant $\gamma$, which essentially meant an augmented conductance for molecules having a higher degree of conjugation [9]. This exponential dependence of SAMs depends on the SAM length ($R_{SAM} = k \cdot e^{(\gamma \cdot length)}$) has been extensively studied and verified. Using a combination of nonequilibrium Green's function, density functional theory and confirmed using reported I-V characteristics, Ratner [5] concluded that the degree of conjugation and overlap of the orbital density of the molecular end group and contact electrode together determine the heightened conductivity of these molecular chains, on account of constant delocalized charge transport. Alkane di-amine molecules have also been studied using common computational chemistry algorithms [10] where the length dependence of electrical conductance has been observed to be exponential, similar to the thiol linker molecules. Generally, most organic molecules demonstrate an exponential dependence of electrical properties on molecule length. However, it must be noted that there is a transition molecular length beyond which the resistance dependence is almost ohmic. Although this critical length depends on the specific organic molecule forming the molecular chain, generally for conjugated molecules, it is ~ 4.0 nm [11]. The exponential dependence of electrical resistance with the length of the molecular chain provides a lower limit of the device resistance; hence for higher resistance change, one should engineer the nanogap and SAM such that the SAM is as short as possible and the air gap should be the same length as the captured gas molecule. This combination maximizes the resistance ratio $R_{OFF}/R_{ON}$ and the dynamic range of the device. The electrical equivalent model for the device is discussed below.

The mechanism for electrical current conduction across the junction of our device is considered to be electron tunneling. In this edge-sensing device shown in the schematic of Figure 3.1 there are two possible electron conduction paths represented by current sources $I_S$ and $I_E(C_g)$, respectively. Current $I_S$ represents electrical conduction through the dielectric spacer stack under the overlapped region of the two electrodes. Current $I_E(C_g)$ represents electrical conduction through the molecular junction and along the edges of the top electrode. This current is a function of the gas concentration $C_g$. In the absence of analyte target gas $C_g = 0$ and since the area under the overlap region occupied by the spacer stack is ~9 $\mu m^2$ and the edge undercut area is ~0.06 $\mu m^2$, current mainly flows through the spacer stack in the overlap region hence $I_E(0) \ll I_S$ thus $I_T \approx I_S$. On the other hand if the target is present, the device gap and linkers should be designed such that $I_E(C_g > 0) \gg I_S$ and $I_T \approx I_E(C_g)$.

Total current flowing across the junction can be written as:

$$I_T = I_S(V) + I_E(V, C_g),$$ (1)

where $I_T$ is the total current flowing across the nanogap sensor and V is the biasing voltage across the device. The substrate and edge current components have different functional forms. The substrate component corresponds to conduction through the thin dielectric stacks is readily modeled using the generalized Simmons' formula for tunneling current [7]. The current across the overlap support region can be expressed as:

$$I_S(V) = G_{SO} \cdot \left[ \frac{\Phi_S}{q} \cdot e^{\left(-2 \cdot d_S \cdot \sqrt{\frac{2m}{\hbar^2} \cdot \Phi_S}\right)} + \left(\frac{\Phi_S}{q} + V\right) \cdot e^{\left(-2 \cdot d_S \cdot \sqrt{\frac{2m}{\hbar^2} \cdot (\Phi_S + qV)}\right)} \right],$$ (2)

where $\Phi_s$ is the barrier energy of the spacer dielectric, $d_s$ is the spacer layer thickness, m is effective mass of tunneling electrons, $\hbar$ is the reduced Planck's constant, q is the charge of electron. The parameter $G_{SO} = A_s \cdot g_{SO}$ is a conductance fitting factor proportional to the area $A_s = W^2$ of the spacer dielectric.

The derivation of a mathematical expression for $I_E$ is considerably more difficult [12, 13], and in general it involves the calculation of nonequilibrium Green's functions which specify the electron transport across the various molecular levels of the trapped gas molecule. Such calculations can lead to tunneling resonances and negative resistance regimes. The model that we have adopted below is based on our experimental observations, which did not display any resonant tunneling characteristics.

In the nontunneling case the *I-V* characteristics resemble those provided by a Simmons' –like model. However, when investigating current conduction across molecular chains, Ghosh [14] found that the Simmons' expression for tunneling current was unable to describe the *I-V* characteristics accurately. The reason is that certain assumptions made during the derivation of the Simmons' expression does not hold true for a molecular chain. However, N. Zimbovskaya [11] found that the conduction through molecular junctions can be modeled by combining a Wentzel-Kramer-Brillouin (WKB) approximation of the transmission coefficient and the Landauer formula which leads to a mathematical expression of tunneling current similar to Simmons' formula. The tunneling current depends exponentially on a function of an average energy barrier as seen by the tunneling electrons. When the analyte gas is captured by our sensor, the dominant current is the gas-dependent edge current through the captured molecules and using Zimbovskaya formula it is given by:

$$I_E(V, C_G) = \left[ \frac{\beta \cdot C_G}{1 + \beta \cdot C_G} \right] \cdot \Gamma_{EO} \cdot \left[ \sqrt{\left( \frac{\Phi_E}{q} + \frac{V}{2} \right)} \cdot e^{-2 \cdot d_E \cdot \sqrt{\frac{2 \cdot m_E}{\hbar^2} \left( \Phi_E + \frac{qV}{2} \right)}} - \sqrt{\left( \frac{\Phi_E}{q} - \frac{V}{2} \right)} \cdot e^{-2 \cdot d_E \cdot \sqrt{\frac{2 \cdot m_E}{\hbar^2} \left( \Phi_E - \frac{qV}{2} \right)}} \right], \quad (3)$$

where, ß is a fitting parameter, $C_g$ is concentration of target analyte gas in the testing chamber, $\Phi_E$ is the average barrier potential of the hybrid molecular chain, and the distance $d_E = 2\times$(length of linker molecule) plus the length of analyte molecule. Two parameters change in this equation when a gas molecule is captured. The first and foremost important gas concentration dependent parameter is the average energy barrier $\Phi_E$ as shown in the zero-bias energy diagram of Figure 3.2. In Figure 3.2, $\Psi_{SAM}$ and $\Psi_{AU}$ are the work-functions of the SAM layer and the gold electrode respectively. $\chi_{target}$ is the electron affinity of the target gas.

Note that the average energy barrier depends on what is inside the edge gap. If there no molecule captured, the average barrier is at the maximum given by the work function of the linker. If a molecule is captured, the average barrier established on that junction depends on the location of the HOMO and LUMO energy levels of the captured molecule. In general, the capture results in a lower average energy barrier $\Phi_E$ hence a large enhancement of the device current.

The second concentration dependent factor is the leading term

$$\Gamma_{E0}\left[\frac{\beta * c_G}{1 + \beta * c_G}\right], \tag{4}$$

which is indicative of a saturation-type Langmuir absorption characteristic. The parameter ß has units of gas concentration and it determines the low-concentration limit behavior of the edge current. This equation implies that the higher the surface concentration of target gas is, the larger the number of molecular junctions formed and the higher the edge current is. The edge current saturates when all possible absorption sites are full. The parameter $\Gamma_{E0} = p \cdot g_{E0}$ is a fitting parameter proportional to the edge perimeter $p=4 \cdot (W+2\Delta)$. This parameter specifies the magnitude of the nonlinear edge conductance and it has units of $A \cdot V^{-1/2}$. Note that Eq. (3) also displays an exponential dependence of the resistance of the molecular junction on total molecular length, which is in agreement with observations made by others [5, 9, 10].

### 3.3.2 Target molecule and SAM capture linker structure

To demonstrate the validity of the proposed sensing mechanism, we designed devices that utilize vapors of 1,5-pentanediamine, that is commonly known as cadaverine as the target gas molecule. Figure 3.3 shows the chemical structure and corresponding length of target and linker molecules. The thiol linker molecule comprises of three components, the sulfur head group, the aromatic and alkyl spacer chain and the benzoic acid tail group. The head group in thiol molecules (-SH) covalently bonds to the gold surface. The spacer chain defines the length of the linker molecule and due to its conjugated nature, allows for augmented electron flow throughout the molecule. The detection of our targets is achieved by the tail segment of the linker, which in this case is composed of benzoic acid. The terminal carboxylic acid, the length of the spacer chain and the nanogap thickness together determine the specificity of the target gas molecules that are to be captured.

### 3.3.3 Electrical characteristics at atmospheric conditions

Figure 3.4 a, b shows I-V characteristics of the fabricated nanogap electrode device in atmospheric conditions and at room temperature. The average DC resistance was found to be ~ 74 GΩ. Measurements indicate that the leakage current across the tunneling junction is in the order of pA until a biasing voltage of about 7.0 V is reached where the spacer dielectric layer instantaneously breaks down. Hence, the leakage power for the nanogap electrode device in its 'OFF' state, is lower than the pW range.

### 3.3.4 Sensor response versus target gas concentration

Sensor response has been previously presented as preliminary results [56]. Figure 3.5 shows the response of the nanogap device when exposed to different concentrations of cadaverine. As evident from the plot, exposure to 30-80 ppm of cadaverine lead to a reduction of junction resistance by ~ 2-8 orders of magnitude respectively. If we define successful switching to demonstrate an ON/OFF ratio of ~two orders of magnitude or more, we can conclude that the threshold for switching our nanogap devices 'ON' is ~30 ppm of analyte concentration in the testing chamber. Although we have observed one order of magnitude change in resistance upon exposure to ~20 ppm of cadaverine, it is too low to be considered as successful switching. Figure 3.6 shows the normalized maximum resistance drop of the nanogap sensor when it is exposed to different concentrations of cadaverine.

### 3.3.5 Experimental and fitted model I-V characteristics at different gas concentrations

Figure 3.7 shows the I-V plots of the nanogap device after successful capture of cadaverine molecules at various concentrations of target gas, curve-fitted with the mathematical model given in Equation (1-4). As evident from the plots, the experimental data is in excellent agreement with the proposed model. Parameter extraction reveals that upon exposure to ~80 ppm of cadaverine gas, the average potential barrier as seen by the tunneling electrons reduces from ~5 eV to ~0.9 eV. This reduction in average potential barrier is responsible for the augmented current flow in the molecular junction since tunneling current varies exponentially with the square-root of the barrier potential. Parameter extraction also revealed that when the device is exposed to ~80 ppm of cadaverine, the fitting parameter β to be 21.685 ppm and $\Gamma_{E0}$ to be 0.3688 A/V1/2 in Equation (3). The maximum root-mean-square-error of the curve-fitting plots shown in Figure 3.7 was found to be ~ 45%, 1% and 0.6% of the average experimental data, for the I-V characteristics of the nanogap junction after exposure to 0 ppm, 60 ppm and 80 ppm of cadaverine respectively.

As mentioned above, after exposure to 80 ppm of cadaverine, the mathematical model reveals a reduction of average barrier potential for tunneling electrons from ~5 eV to ~0.9 eV. It must be noted that this value is not a quantitative measurement of the HOMO-LUMO levels of individual target/linker molecules but a cumulative indication of the average potential barrier faced by tunneling electrons. In fact Zimbovskaya [11] mentions that the inherent disadvantage of this mathematical model is that although one can accurately describe I-V characteristics using Equation 3, detailed information including the electrostatic potential profile of the transport channel cannot be obtained. This is because of the WKB approximation for electron transmission functions used to derive the model in the first place.

### 3.3.6 Selectivity of sensor response

To investigate the cross-sensitivity of the sensor, we exposed the device to commonly encountered VOCs like acetone, ethanol and hexane as well as gases such as Helium, Nitrogen and $CO_2$. We define the sensor response as the normalized junction resistance drop after exposure to the analyte. Figure 3.8 shows the resistance ratio of the nanogap sensor when exposed to these analytes as compared to the ratio when exposed to the intended target gas, cadaverine. Measurements reveal a $R_{OFF}/R_{ON}$ ratio of more than four orders of magnitude when exposed to only 40 ppm concentration of cadaverine whereas a maximum $R_{OFF}/R_{ON}$ ratio of ~two orders of magnitude, when exposed to much higher concentrations of the other analytes. The concentration levels of the VOCs was maintained at greater than 10,000 ppm. To measure the device response in presence of other gases, we flooded the test chamber with the specific test gas and then monitored the resistance drop of the sensor. These results suggest a highly selective sensor action against most commonly found VOCs.

Figure 3.9 shows resistance change across the molecular junction over one complete cycle of exposure to 40 ppm concentration of target gas and its subsequent removal from gas testing chamber. The plot also shows control (reference) chip signal (a device which was not coated with the linker molecule and exposed to similar ppm amounts of target cadaverine gas). The feeble response of the control chip shows that in absence of proper functionalization, even after exposure to cadaverine gas, there is almost no sensor response as the thiol molecules are unable to capture the analyte molecules and the molecular switch remains 'OFF'. Essentially, the sensor response from the control chip proves the validity of the linker molecules and their essential role in capturing the target gas molecules.

### 3.3.7 Adsorption-desorption dynamics

Amines are known to be notoriously "sticky" compounds [15, 16], which is why it is difficult to analyze them at low concentrations using even sophisticated GC-MS systems. These gases have a tendency to stick to walls of delivery tubes or testing chambers, which makes it extremely challenging to detect under normal conditions. Once these gases adsorb on a given surface, they tend to stick on unless forcefully desorbed using heat treatments or appropriate cleaning protocols. Similarly, we observed that after the analyte was trapped within the nanogap junction, un-assisted and complete recovery was not repeatable. A fraction of the captured analyte always remained trapped within the device and with progressive analyte exposure – removal cycles, the remnant cadaverine molecules kept increasing till device response was negligible. Figure 3.10 shows normalized junction resistance versus time upon successive cycles of device exposure to cadaverine gas. As evident from the plot, every progressive switching cycle displays a reducing baseline resistance drift. This reduction in junction resistance is the result of remnant cadaverine molecules remaining trapped in the molecular junction and contributing to tunneling current even in the absence of exposure to additional cadaverine molecules.

The dynamics of junction resistance drop in the sensor is governed by a quasi-irreversible process of cadaverine molecules forming hydrogen-bonds with the linker end group and forming a hybrid molecular bridge. This is typical of a chemisorption process which has been previously described mathematically by Elovich's equation [17]. The desorption process ideally consists of complete breaking of hydrogen bonds between the target gas molecules and the linker end-group. This typically is characteristic of a simple first-order desorption process where the adsorbed gas-molecule on the surface of the solid simply desorbs back into a gaseous form. The kinetic process is mathematically described by the Polanyi-Wigner equation and is used as the theoretical basis for thermal desorption spectroscopy process [18]. Figure 3.11 shows the normalized conductance versus time plot of one complete cycle of analyte exposure to the device and its subsequent removal, curve-fitted with the Elovich equation for the adsorption cycle and Polanyi-Wigner equation for the desorption cycle. The plot suggests that the mathematical model is in decent agreement with the experimental data. The average root-mean-square-error of the curve-fitting is ~48% of the average data.

## 3.4 Conclusions

We present a new class of chemiresistors based on the capture of gas molecules within a nanoscale gap. We fabricated gas-sensing devices with gold electrodes separated by a ~6 nm nanogap functionalized with a fully conjugated ter-phenyl linker molecule, (4-((4-((4-mercaptophenyl)ethynyl)phenyl)ethynyl)benzoic acid) for electrostatic capture of cadaverine. We demonstrated ultra-low power resistance switching in batch-fabricated nanogap junctions upon detection of target analyte - cadaverine. The stand-by power consumption was measured to be less than 15.0 pW and the $R_{OFF}/R_{ON}$ ratio was more than eight orders of magnitude when exposed to ~80 ppm of cadaverine. A phenomenological electrical model of the device is also presented in good agreement with experimental observations. Cross-sensitivity of the gas sensor was tested by exposing the device to some of the commonly found VOCs and other atmospheric gases. The experiments revealed a highly selective sensor action against most of these analytes. These batch-fabricated sensors consume ultra-low power and demonstrate high selectivity; therefore, they can be suitable candidates for sensor applications in power-critical IoT applications and low power sensing.

## 3.5 <u>References</u>

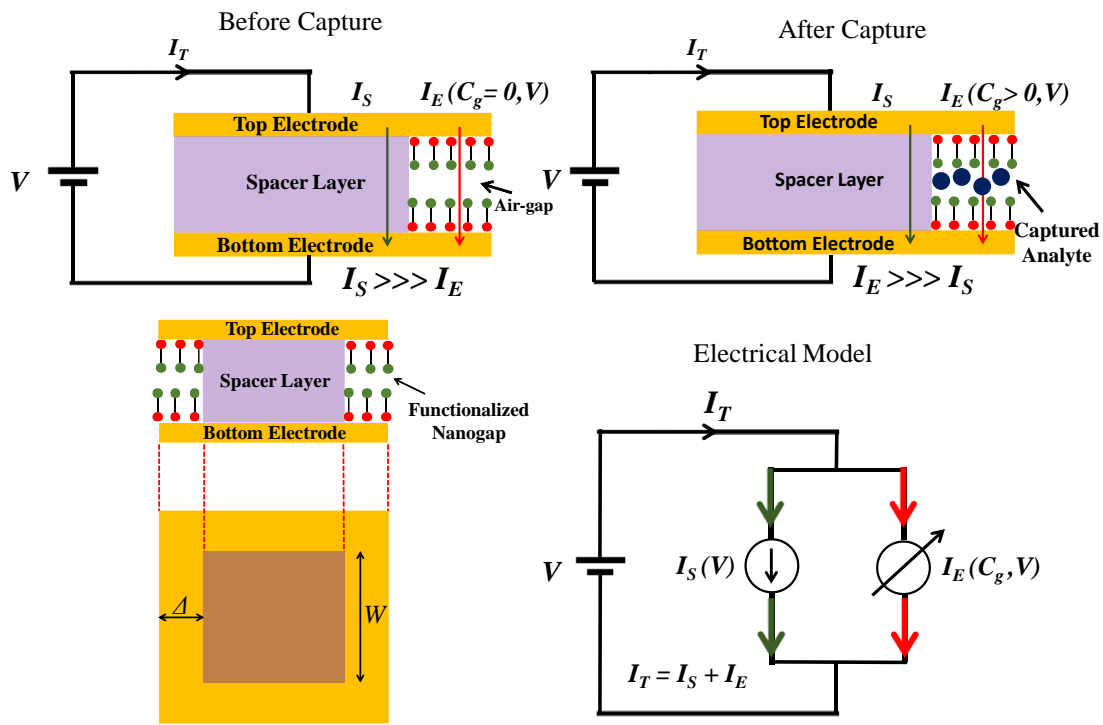

**Figure 3.1**: Schematic representation of current conduction before and after analyte gas capture and equivalent electrical model depicting the two current conducting paths.

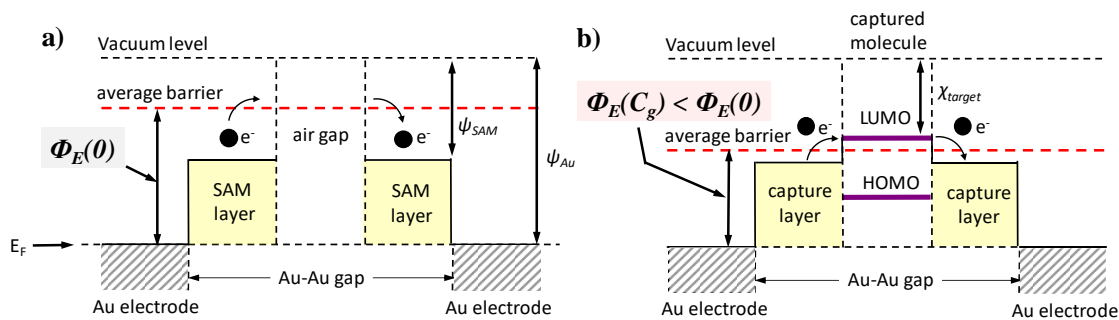

**Figure 3.2**: Barrier potential modification due to molecule capture. a) Average edge energy barrier in absence of target analyte gas. b) Average edge energy barrier at a molecular junction established by the capture of a target gas molecule. The average energy barrier is lowered leading to a larger tunneling current.

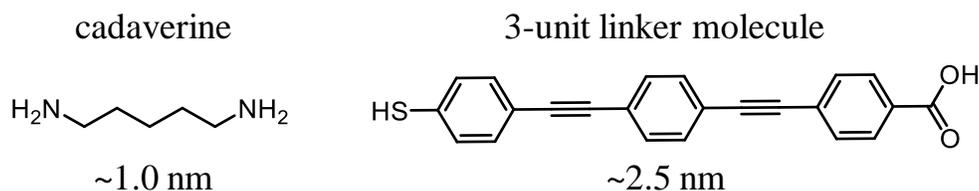

**Figure 3.3**: Chemical structure and molecule-length of the target molecule (on the left) and linker molecule (on the right). The IUPAC name of the linker molecule is (4-((4-((4-mercaptophenyl)ethynyl)phenyl)ethynyl)benzoic acid).

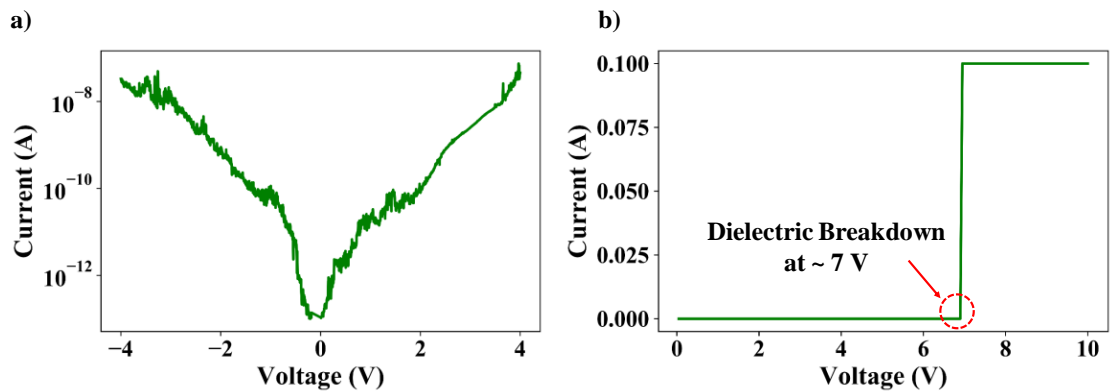

**Figure 3.4**: Electrical characteristics of device. (a) I-V curve of the completed device showing complete electrical isolation of the upper and lower gold electrodes. (b) I-V characteristics showing instantaneous breakdown of dielectric layer at ~ 7 V.

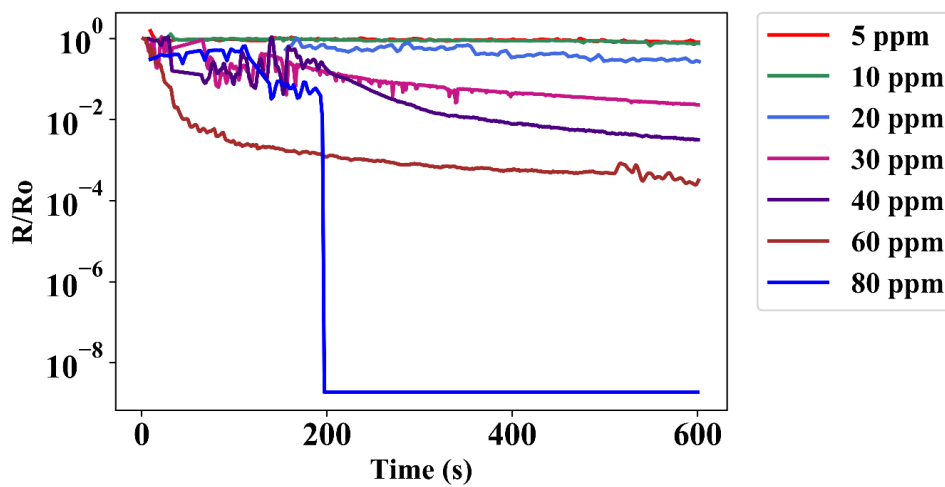

**Figure 3.5**: Response of the functionalized nanogap sensor when exposed to different concentrations of cadaverine target analyte.

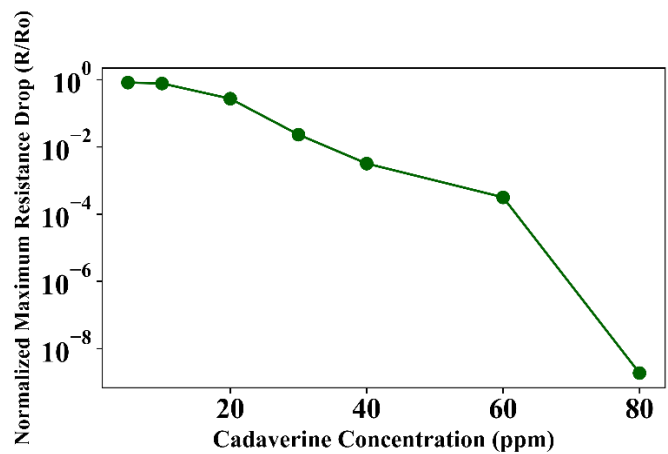

**Figure 3.6**: Normalized maximum resistance drop of nanogap junction after exposure to different concentrations of cadaverine gas.

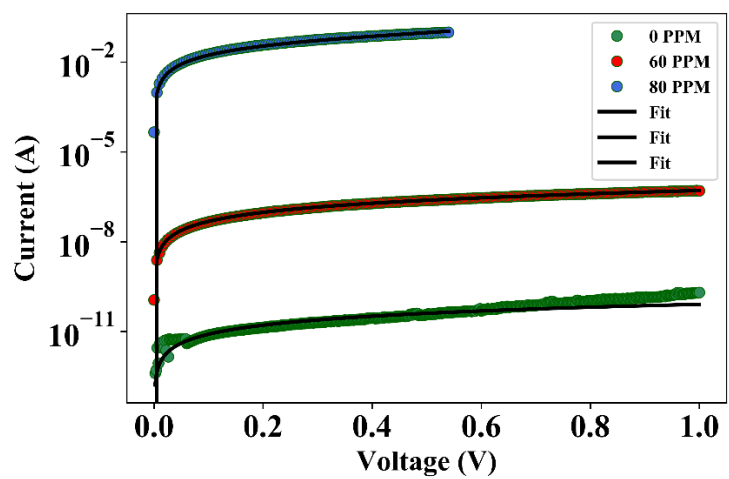

**Figure 3.7**: I-V measurements of the nanogap device after successful capture of cadaverine molecules at various concentrations of analyte.

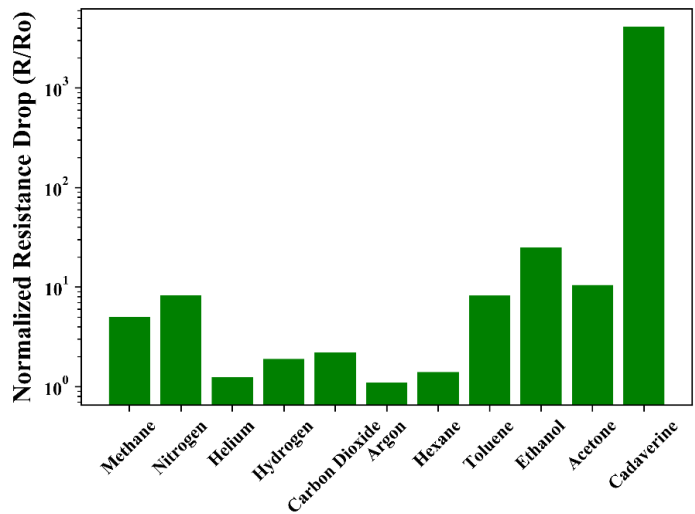

**Figure 3.8**: Device response to commonly found VOC's and other gases.

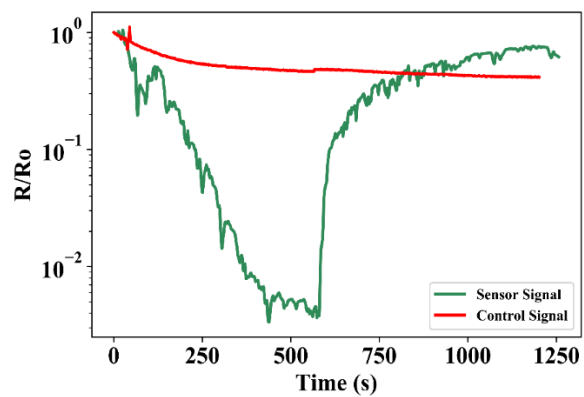

**Figure 3.9**: Sensor analysis over one complete cycle of target gas exposure and its removal. The sensor signal is compared to the response of a control chip which is our nanogap device in absence of proper functionalization.

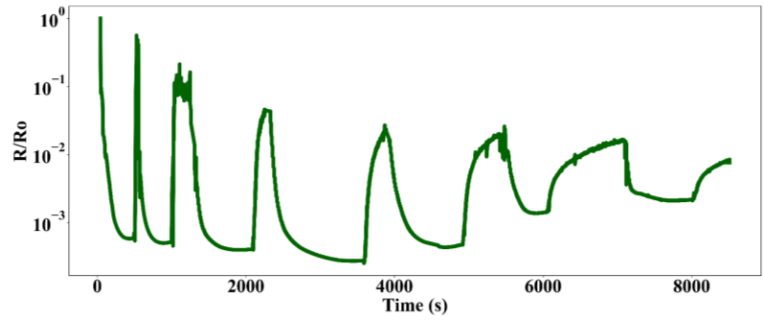

**Figure 3.10:** Repeatability analysis of nanogap sensor.

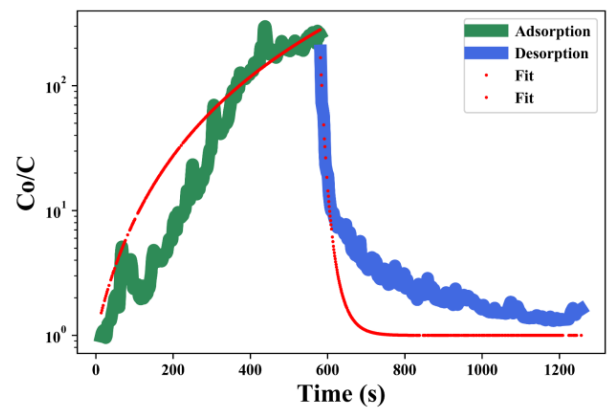

**Figure 3.11:** Adsorption and desorption dynamics of the sensor action.

# CHAPTER 4

# ELECTRICAL DETECTION OF PROTEINS USING BATCH-FABRICATED VERTICAL METAL NANOGAP BREAK-JUNCTIONS

## 4.1 Introduction

In this chapter, using the fabrication scheme explained previously in Chapter 2, we batch-fabricate nanogap electrodes and use them for label-free protein detection.

The specific detection of proteins and nucleic acids is necessary for understanding a plethora of biological phenomena. Concentration dependent interactions have been detected using labeled and label-free methods such as laser-induced fluorescence (LIF), surface plasmon resonance (SPR), amperometry and mass spectrometry [1]. The interactions can also be detected from changes in the electrical conductivity in metal-molecule-metal nanometer bridge junctions [1-3]. In this paper we report the fabrication and implementation of a vertical nanogap device for conductivity-based detection of proteins. The vertical nanogap devices are batch-fabricated using conventional planar processing, offering significant advantages over nanogap devices fabricated using e-beam lithography, electrodeposition and focused-ion-beams (FIB) [1]. Figure 4.1 shows the working principle behind electrical detection of bio-molecules in nanogap junctions.

## 4.2 <u>Experimental Procedures</u>

### 4.2.1 Device design, structure, and fabrication

Figure 4.2 shows a schematic of the vertical nanogap device. For good insulation between the nanogap electrodes withstanding more than 10 V, we have implemented a vertical nanogap realization with the spacer layer sandwiched between two gold electrodes. The device includes a thin film of >10 nm of silica ($SiO_2$) is vacuum sputtered on a 500 nm thick gold (Au) film followed by >3 nm film of silicon (Si). We have implemented two such devices using total thicknesses of 15 and 50 nm. The thin $SiO_2$ prevents tunneling current between Au electrodes and the Si film offers a good adhesion for the upper electrode of 300 nm thick Au film. After patterning rectangular patches of the film stack the Si and $SiO_2$ are notched along the top electrode edge thus forming nanogaps along its perimeter. Figure 4.3 shows the vertical nanogap device fabrication process flow. Figure 4.4 shows scanning electron microscope images of the device.

### 4.2.2 Device functionalization and sample preparation

Top and bottom Au electrode surfaces are first functionalized with a dual-PEG brush of long-chain 5k Da carboxy-PEG-thiol and short-chain 2k Da methoxy-PEG-Thiol [4] for 50 nm device and 1k Da carboxy-PEG-thiol for 15 nm device. This configuration ensures elimination of nonspecific binding of proteins via excellent antifouling characteristics of poly (ethylene glycol) (PEG) and the anchoring of proteins in a sandwiched format between the PEG chains. For protein anchor demonstration, we utilize well-established EDC/NHS cross linking chemistry between the carboxy head group and amino groups of surface lysine residues of Bovine serum albumin (BSA) and Bovine Carbonic Anhydrase II (CA-II).

## 4.3 <u>Results</u>

Figure 4.5 shows the I-V characteristics of the junction after device fabrication. The plots suggest an extremely low leakage current under normal biasing conditions. Figure 4.6 shows the I-V characteristics of the device after successful functionalization. The plot suggests that even after the PEG molecules have formed a self-assembled layer on the bare gold surface, it hasn't bridged the nanogap junction. A linear I-V characteristic is observed only for the final step of protein anchoring, as shown in Figure 4.7. The plots shown in Figure 4.8 indicate that the nanogap resistance decreases for longer periods of cross-linking. After exposure to the proteins, the junction resistance of the nanogap device decreased by nine orders of magnitude.

## 4.4 <u>Conclusion</u>

We believe that this device has the potential for high-yield low-cost performance of protein concentration measurements, especially for on-site and point-of-care applications. The nanogap dimensions and the chain length of PEG brushes can be tuned to address the specific sizes of medically relevant proteins and other bio-molecules, thus adding to the versatility of electrical conductivity experiments for a wide range of bio-samples

## 4.5 <u>References</u>

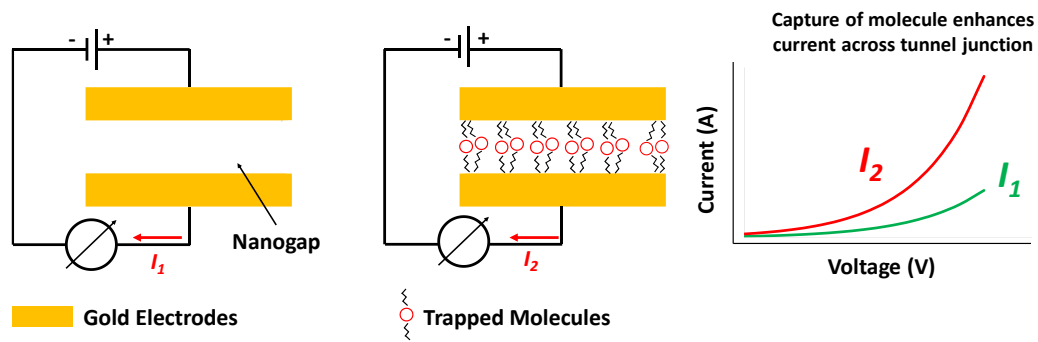

**Figure 4.1**: Schematic of working of nanogap biosensors. Trapping of bio-molecules within the nanogap leads to augmented current across junction.

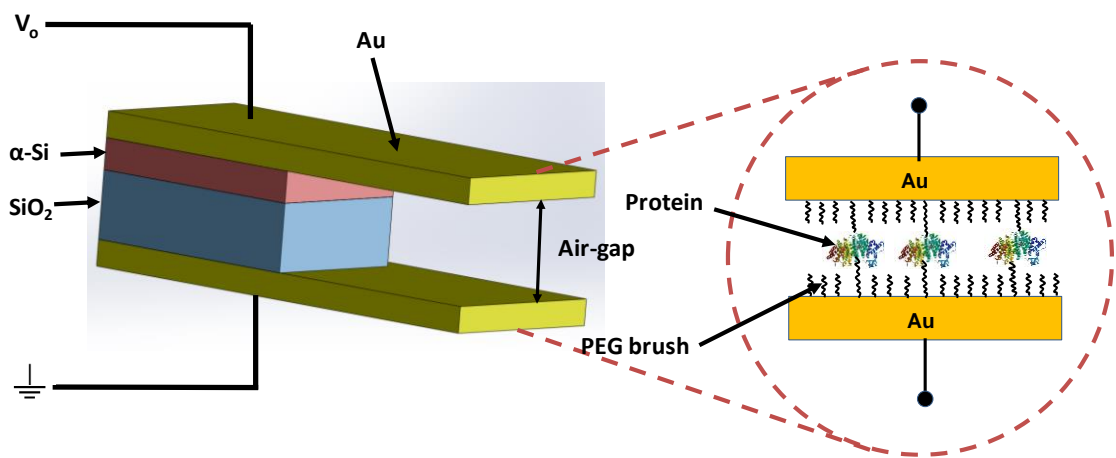

**Figure 4.2**: Schematic of fabricated device. The protein is specifically captured within the nanogap junction by the PEG brushes. The electrical conductivity of the device increases by nine orders of magnitude when the protein is captured.

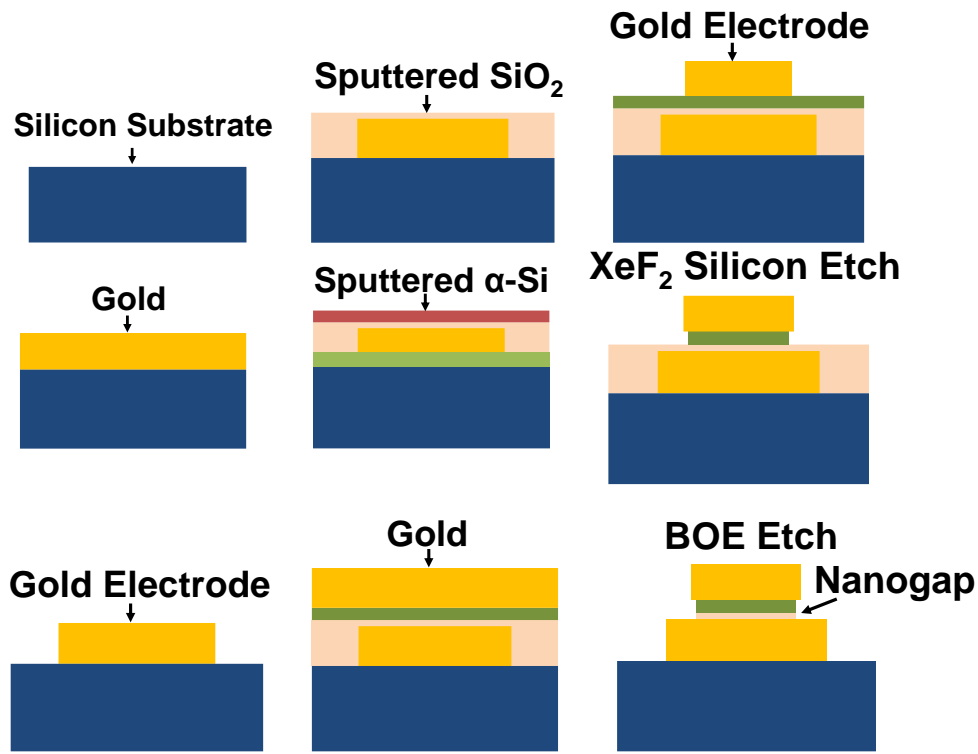

**Figure 4.3**: Fabrication flow for the nanogap bio-sensor. The combined thickness of the α-Si and sputtered SiO2 together define the spacer gap.

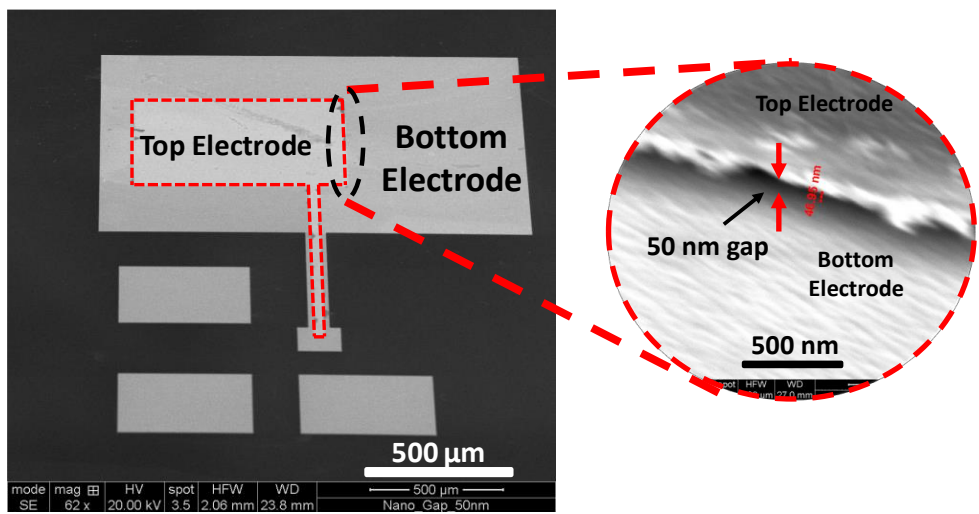

**Figure 4.4**: SEM image of fabricated device. The insert on the right shows a zoomed in view of the 50 nm air-gap between top and bottom electrodes.

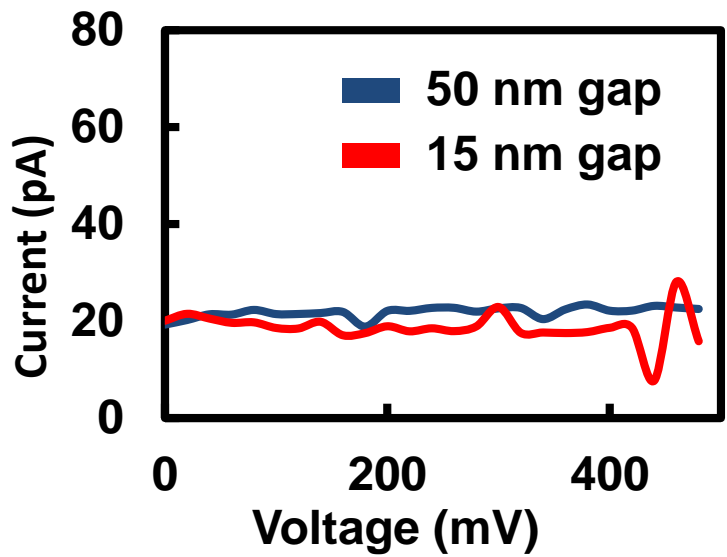

**Figure 4.5**: I-V characteristics of the nanogap junction after device fabrication for nanogap devices having both 15 nm and 50 nm spacer thicknesses.

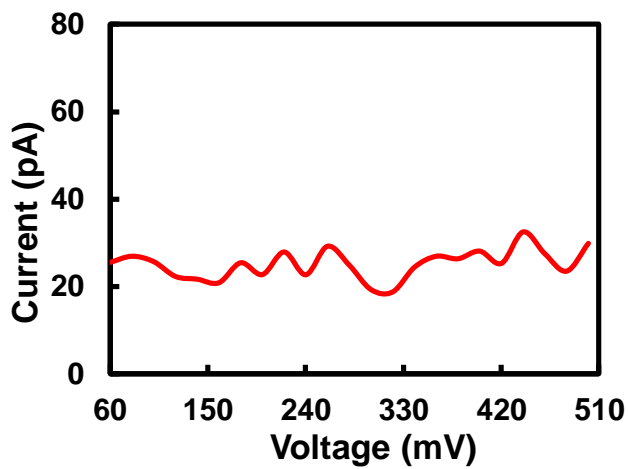

**Figure 4.6**: I-V characteristics of the nanogap junction after functionalization using 5k Da carboxy-PEG-thiol.

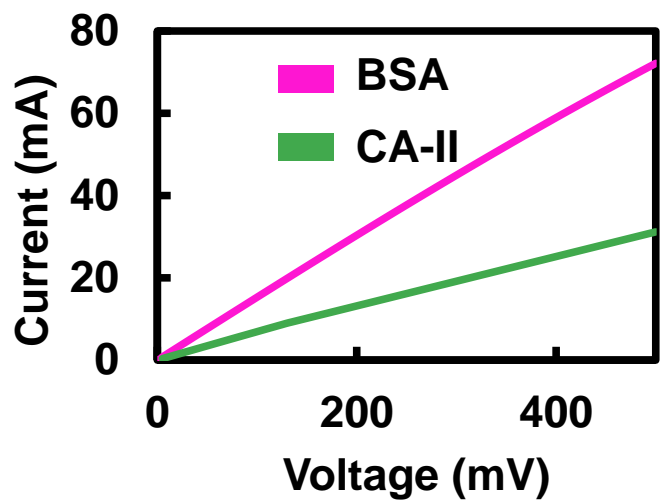

**Figure 4.7**: I-V characteristics of the nanogap junction immersion in bovine serum albumin (BSA) and bovine carbonic anhydrase II (CA-II).

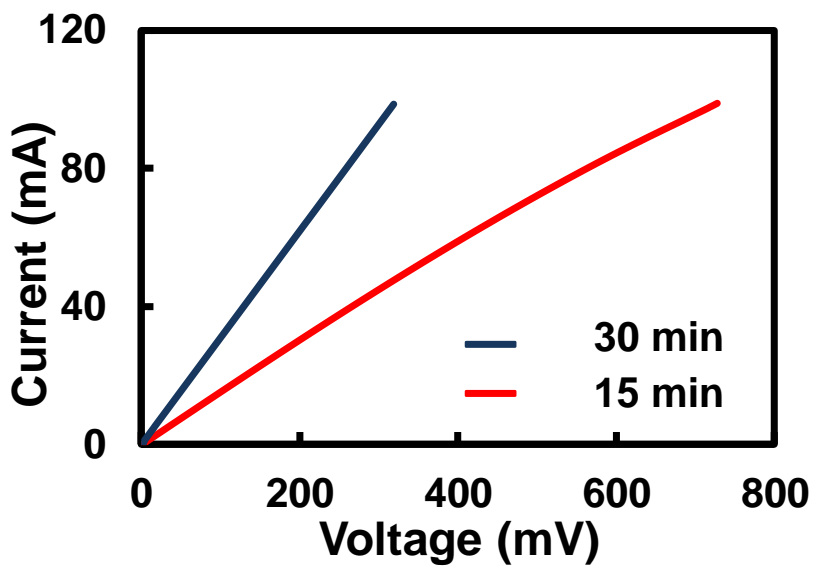

**Figure 4.8**: I-V characteristics of the nanogap junction as a function of immersion time in bovine serum albumin (BSA).

# CHAPTER 5

# QUANTUM TUNNELING HYGROMETER WITH TEMPERATURE STABILIZED NANOMETER GAP

## 5.1 Introduction

Humidity sensors can be realized using various technologies [1-5]. Kharaz and Jones were one of the first to report optical methods of humidity measurement [6]. First introduced by Wohltjen in 1984, SAW sensors were widely used as hygrometers, where shift in resonant frequencies was a direct indication of mass of adsorbed water vapor [7]. The two most common types of hygrometer devices are resistive and capacitive type.

Capacitive humidity sensors are the most common realizations of microfabricated hygrometers [8]. These devices are used in approximately 75% of applications [3]. In these devices, the dielectric properties of a hygroscopic layer (usually a polymer such as polyimide) change when exposed to humidity. This change may also be accompanied by swelling of the hygroscopic layer. Microcantilevers have also been used as capacitive type humidity sensors. In these devices, a thin polyimide film is patterned on the cantilever and when exposed to humidity, the polymer film tends to expand thereby exerting mechanical stress on the cantilever causing it to bend [9]. Capacitive humidity sensors display linear operation and can accurately measure from 0% -100% relative humidity with an output range typically ~30% of the default capacitance. These devices consume very little electrical power and are relatively cheap to fabricate[3].

Resistive humidity sensors are also extensively used. In these devices the resistance of the sensing material changes when exposed to humidity. Ceramics such as $Al_2O_3$ [10], conductive polymers such as PVA with graphitized carbon [11] and polyelectrolytes [12] have been used as functional materials for resistive humidity sensors. In contrast to capacitive devices, resistive humidity sensors are highly nonlinear and capable of providing several orders of magnitude change in resistance. An advantage of resistive RH sensors is their interchangeability, typically within ±2% RH. Operating temperatures of resistive sensors range from −40°C to 100°C; however these devices have significant temperature dependencies. For example, a typical temperature coefficient for a resistive RH sensor is -1% RH / °C.

In this chapter we present a new type of microfabricated humidity sensor that can provide large output range and low temperature dependence. The device utilizes the expansion of a polymer that swells when exposed to humidity as in the capacitive device, but it produces a resistive output that measures the polymer expansion through tunneling current across a humidity dependent, thermally stabilized nanogap. The tunneling current changes many orders of magnitude providing similar output as the resistive type device with a low temperature dependence.

### 5.2 <u>Basic Operating Principle of Resistive Nanogap Hygrometer</u>

The device consists of a pair of upper and lower electrodes, separated by an air-gap of ~2.5 nm as shown in Figure 5.1(a) below illustrating the basic sense mechanism.

The upper electrode rests at a fixed height. The bottom electrode rests on top of a hygroscopic polymer, polyimide that swells when humidity is absorbed; thus reducing the gap between the two electrodes. The swelling of the polyimide with humidity is linear corresponding to its humidity coefficient of expansion or CHE, which is 60-75 ppm/% RH [13]. Figure 5.1(b) shows the electron band diagram across the nanogap. If the nanogap is very small electrons can tunnel from one electrode to the other establishing a conductive path. The magnitude of the tunneling current is exponentially dependent on the gap [14] hence the device resistance goes as

$$R(RH\%) = R_o \cdot e^{(a*\Delta(RH\%))} \sim A(T) \cdot e^{(-B \cdot (RH\%))}, \qquad (1)$$

where A(T) and B are fitting parameters. In the simple configuration shown in Figure 5.1(a), the gap $\Delta$ is also affected by the thermal expansion of the polyimide, which makes the coefficient A and the resistance of the device strongly dependent of the ambient temperature.

To eliminate this strong temperature dependence we utilize the differential device arrangement as shown in the schematic of Figure 5.2 where the two electrodes rest on polyimide patches of equal thickness, hence in the absence of humidity the nanogap distance is constant and independent of temperature fluctuations. However only one of these patches can absorb humidity; thus producing a humidity-induced nanogap change. The differential device assembly consists of an upper Al electrode and a lower Cr electrode separated by an air-gap of ~2.5 nm, standing on separate patches of 1.5 µm thick polyimide. As shown in Figure 5.2, the polyimide patch under the upper electrode is covered with ~8 nm of ALD deposited $Al_2O_3$ diffusion barrier whereas that under the lower electrode is exposed to ambient atmosphere.

We thus achieve nanogap temperature stabilization by using cancelation of a common mode thermal expansion of both patches and humidity signal extraction by differential response to humidity between the two patches. Since both the top and bottom electrodes are standing on near identical polyimide patches, any increase in ambient temperature would lead to both the patches expanding almost equally. This ensures that in the event of temperature fluctuation, the nanogap distance between the top and bottom electrode will remain unchanged and the electrical response will be negligible in comparison to sensor response.

## 5.3 Experimental Procedures

### 5.3.1 Device fabrication

The fabrication process started by growing ~1 µm of thermal $SiO_2$ on a 4-inch Si wafer. Polyimide was then diluted by dissolving uncured HD4104 polyimide (purchased from HD Microsystems) in N-Methyl-2-pyrrolidone (NMP) solvent in the ratio of 1:0.5 (wt/wt). This mixture was then dissolved by using a stirring it with a magnetic stirrer at 300 rpm for 2 hours to ensure perfect mixing of the polyimide and the NMP solvent. This mixture was then spin-coated on the sample at 2000 rpm following the standard procedures for polyimide processing. For curing the polyimide, the sample was kept in a nitrogenized environment oven for three hours at a temperature of 300 ° C. This procedure resulted in polyimide thickness of ~1.5 µm. Following this, we sputtered 200 nm of Al on the sample at 50 W and used it as a hard-mask to pattern the underlying polyimide. $O_2$ plasma dry etching for 10 minutes at 100W was sufficient to remove the unwanted polyimide from the sample. Next, the remnant Al was stripped off by using commercially available aluminum etchant. Following this, we sputtered ~ 100 nm of Cr at 50 W on the sample and lithographically patterned it to define the lower electrodes. After this, we deposited about 2.5 µm of PECVD α-Si on the sample to form the sacrificial bridging supports for the upper electrode. This was followed by thermal ALD of ~8 nm $Al_2O_3$ and its lithographic patterning using BOE. Then, we sputtered ~2.5 nm of α-Si on the sample at 50 W to pattern the sacrificial spacer layer to define the thickness of the eventual nanogap between the electrodes. We next deposited ~1 μm of Al on the sample and patterned it to form the upper electrode. Finally, we sacrificially etched away the α-Si using $XeF_2$. A total of 1200 minutes of etching was required to completely etch away the sacrificial Si and release the upper electrode.

### 5.3.2 Imaging, electrical characterization, and sensor response

High-resolution SEM imaging was done at an accelerating voltage of 15.0 kV by the FEI Nova NanoSEM to visually inspect the fabricated device. I-V characteristics of the device were measured using the Keithley 4200A-SCS Semiconductor Parameter Analyzer. Tunneling current measurements were performed in a dark room, where the device under test was placed on a probe-station kept inside an electrically shielded enclosure to ensure low-noise and high fidelity electrical signals. The enclosure had feedthroughs for providing electrical connections to the probe-station. An inlet from a commercially available humidifier with restricted flow was used to control and maintain the humidity levels in the enclosure. An Arduino powered BME280 chip was placed inside the chamber for serial monitoring of the relative humidity levels.

### 5.4 Results and Discussion

### 5.4.1 Fabrication and electrical characterization

The simplified fabrication process is shown in Figure 5.3. A more detailed description of the fabrication process is given in the "methods" section. Figure 5.4 a shows high resolution SEM images of the fabricated sensor and Figure 5.4 b shows *I-V* characteristics of the device after fabrication of the device. The junction *I-V* characteristics are typical of very low current on account of very low electron tunneling across an air-gap of ~2.5 nm. The average DC resistance of the junction is measured to be ~271 GΩ. This shows an extremely low leakage current under low biasing conditions.

### 5.4.2 Sensor action and tunneling current at various relative humidity levels

After the device is fabricated, the upper and lower electrodes are separated by an air-gap of ~ 2.5 nm. Therefore current flow across the nanogap junction involves conducting electrons tunneling through 2.5 nm of air-gap. However, when the device is exposed to an increase in humidity, only the un-protected patch of polyimide which is beneath the lower electrode absorbs water-vapor molecules and swells. This is because the $Al_2O_3$ acts as a diffusion barrier and prevents the polyimide patch under the upper electrode from absorbing water molecules [15]. This differential swelling of the polyimide patches results in the inter-electrode distance to reduce below 2.5 nm. Therefore, after absorption of water vapor molecules, the tunneling distance for the conduction electrons reduces. Since tunneling current exhibits exponential dependence on tunneling distance, increase in humidity levels lead to an exponential increase in tunneling current. In other words, junction resistance decreases exponentially when exposed to increased humidity levels.

Figure 5.5 a shows the I-V curves of the device when exposed to an increasing RH% from ~20 − 90 RH%. As evident from the *I-V* plots, the increasing RH% leads to an increase in magnitude of current flow for the same voltage bias. Figure 5.5 b is a plot of normalized average resistance vs RH%. These plots are a clear indication of exponential reduction of junction resistance with increasing RH%. Figure 5.6 a-c shows five cycles of repetitive exposure of device to an increase in ambient RH% from 20-90 for three separate devices and subsequent removal of excess humidity from the testing chamber. As evident from the plot, the device exhibits unassisted and perfect recovery of junction characteristics after removal of water vapor from the testing chamber. The plots also suggest that there is a difference in time taken to reach the maximum resistance drop for different devices. The reason is simply because of the fact that the time taken for the chamber to reach the same humidity level varied between successive testing cycles and during multiple device testing. Our fabricated device followed the commercially available BME280 reference sensor perfectly. Any delay in sensor response was also exhibited by the reference chip as well. Figure 5.7 shows the comparative plots of successive cycles of moisture absorption and desorption by the device. As evident from the plot, the device doesn't suffer from moisture hysteresis or sensor saturation.

### 5.4.3 Temperature response of the device and passive temperature compensation

Researchers have always tried to neutralize the undesired temperature response of polymer-based sensors [9]. Essentially, temperature compensation can be achieved using passive and active methods. In passive temperature compensation, passive electrical elements such as resistors are employed to compensate predetermined fluctuations in sensor readout due to temperature changes. Active temperature compensation involves active temperature feedback to the transducer's signal processing circuit to compensate for the temperature drift. Recently, Rinaldi [16] and Mastrangelo [17] used self-levelling beams in order to achieve temperature compensation in micromachined sensors. We have previously shown very preliminary results for our quantum-tunneling assisted hygrometer [21,22]. In our proposed tunneling hygrometer, we achieve passive temperature compensation by using a nondifferential polymer expansion design when exposed to temperature fluctuations. As shown in Figure 5.8, when the device is exposed to water vapor molecules, the polymer platform which does not have a protective layer of $Al_2O_3$ absorbs the molecules and expands, but the protected polymer patch does not. Therefore, air-gap between the electrodes reduces and tunneling current flow between them increases. However, when the device is subjected to an increase in temperature, due to near identical thermal mass and negligible thickness of the $Al_2O_3$ layer in comparison with the platform, both polyimide patches expand equally. This ensures that the air-gap between the electrodes remains almost constant.

Figure 5.9 shows the temperature response of the device when subjected to a temperature change from 25-60 °C. As evident from the plot, the junction reduces <5 times when subjected to temperature changes. This is ~0.05% of the maximum resistance drop of the junction resistance when exposed to change in humidity. Therefore, it is clear that the device displays sufficient temperature compensation.

### 5.4.4 Current conduction mechanism, mathematical model, and equivalent electrical circuit

The current conduction through the nanogap junction is considered to be electron tunneling. Essentially, current flow across the air-gap is possible only when conduction electrons tunnel between the metal electrodes. As explained above, when the device is exposed to an increasing RH%, the gap between the electrodes reduce. Tunneling current exponentially depends on the distance through which the electrons have to tunnel through. Therefore, exposure to rising levels of humidity leads to an exponential reduction of junction resistance. The air-gap between the upper and lower electrode can then be considered as a function of RH% and coefficient of hygroscopic expansion of polyimide. The current conduction can be modeled by using the generalized Simmons' expression for electron tunneling current density [18]. Figure 5.10 a shows the equivalent electrical circuit describing our sensor action.

For intermediate voltages, the current flowing through the nanogap junction can be written as:

$$I_T(V, C_g) = (G_{SO}) \cdot \left[ \frac{\Phi_a}{q} \cdot e^{\left( -2 \cdot ((d_s - \beta \cdot t_p \cdot C_g) \cdot \sqrt{\frac{2m}{\hbar^2} \cdot \Phi_a}) \right)} + \left( \frac{\Phi_a}{q} + V \right) \cdot e^{\left( -2 \cdot (d_s - \beta \cdot t_p \cdot C_g) \cdot \sqrt{\frac{2m}{\hbar^2} \cdot (\Phi_a + qV)} \right)} \right], \qquad (2)$$

where, $I_T$ (V, $C_g$) = current across the air-gap as a function of applied voltage and relative humidity of the testing chamber, V = Biasing voltage across device, $C_g$ = Relative humidity of the testing chamber, ß is a fitting parameter proportional to the coefficient of hygroscopic expansion for polyimide, $t_p$ = thickness of polyimide platform, $\Phi_a$ is the mean barrier potential of the air-gap , $d_s$ = original air-gap, m = effective mass of tunneling electrons, ħ = reduced Planck's constant, q = charge of electron and $G_{SO}$ is a conductance fitting factor proportional to the overlap area and having units of S.

Figure 5.10 b shows the I-V plots of the device after being exposed to varying levels of relative humidity curve fitted with Equation 1. As evident from the plot, the experimental data is in good agreement with the mathematical model. Parameter extraction revealed the average value of ß to be ~2.78 ppm/RH% and $G_{SO}$ to vary from $5.2 \times 10^{-4}$ $\Omega^{-1}$ for the I-V curve corresponding to 24 RH% to ~ 10 $\Omega^{-1}$ for the I-V curve corresponding to 83 RH%. The root-mean-square-error of the curve-fitting plots shown in Figure 10b was found to be 10%, 13%, 1.5% and 6% of the average experimental data of the I-V curves corresponding to junction electrical characteristics when the sensor was exposed to 24 RH%, 44 RH%, 62 RH% and 83 RH%, respectively.

### 5.4.5 Absorption-desorption dynamics

Since sensor dynamics is heavily dependent on the diffusion of water vapor molecules into the polyimide patch causing expansion of the polymer, the dynamic response of the absorption cycle of the sensor can therefore by mathematically described using a modified version of Fick's second law [19]. Fick's law of diffusion has been extensively used to describe the diffusion process and to experimentally determine diffusion constant of various substances. The desorption process ideally consists of excess water molecules in the polyimide desorbing back into the atmosphere into water-vapor molecules. This typically is characteristic of a simple first-order desorption process where the absorbed gas-molecule on the surface of the solid simply desorbs back into its gaseous form. The kinetic process is mathematically described by the Polanyi-Wigner equation and is used as the theoretical basis for thermal desorption spectroscopy process [20]. Figure 5.11 shows the normalized conductance of the sensor as a function of time for one cycle of water-vapor absorption and desorption, where the absorption cycle is curve-fitted with Fick's law of diffusion and the desorption cycle is curve-fitted with the Polanyi-Wigner desorption model. As evident from the plot, the presented mathematical model accurately describes the sensor action.

### 5.5 <u>Conclusions</u>

We presented the design, fabrication, electrical characterization and working of a temperature compensated tunneling humidity sensor. Sensor response shows a completely reversible reduction of junction resistance of ~five orders of magnitude with a standby DC power consumption of ~0.4 pW when the device is exposed to an increasing RH% of ~ 20-90 RH%. Passive temperature compensation was also achieved using a nondifferential swelling design for the polymer patches. Temperature response showed a reduction of ~2.5 times when the device was exposed to a temperature sweep of 25°C – 60°C, which is 0.0025% of the maximum sensor output when exposed to rising levels of humidity. A mathematical model and equivalent electrical circuit were also presented to accurately describe the current conduction mechanism responsible for sensor action. Finally, sensor response dynamics was also investigated and established sorption analytical models were used to describe the time-dependent sensor action.

## 5.6 <u>References</u>

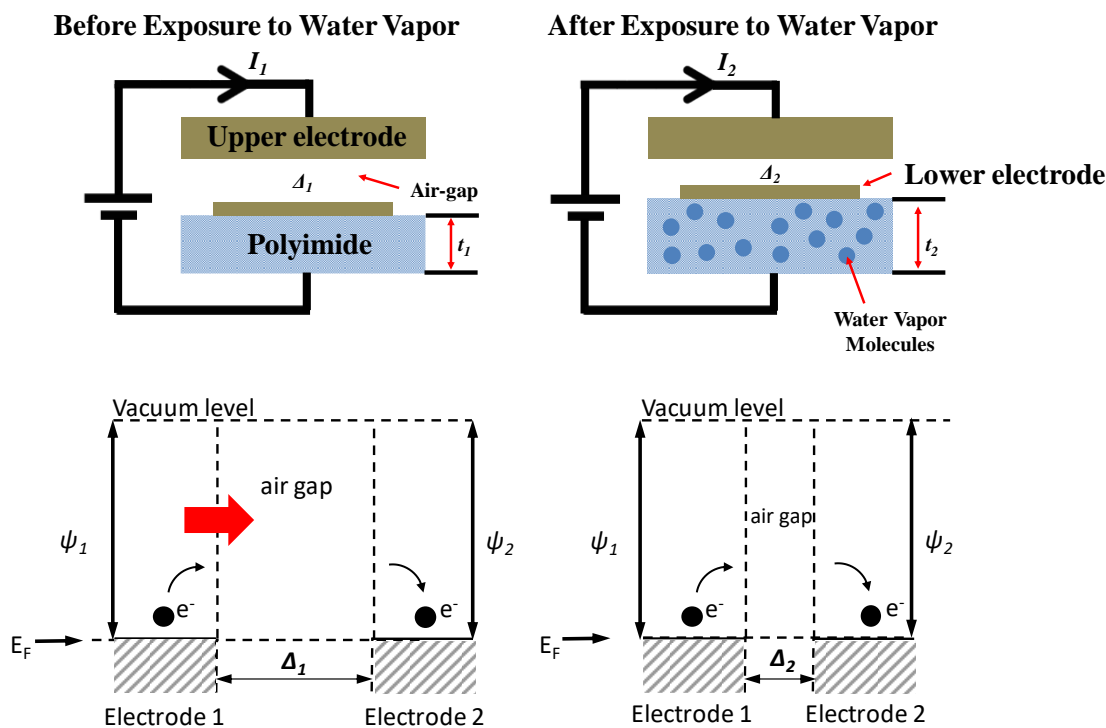

**Figure 5.1**: Working principle of the tunneling humidity sensor. Exposure to water vapor molecules reduces the gap and reduces tunneling distance.

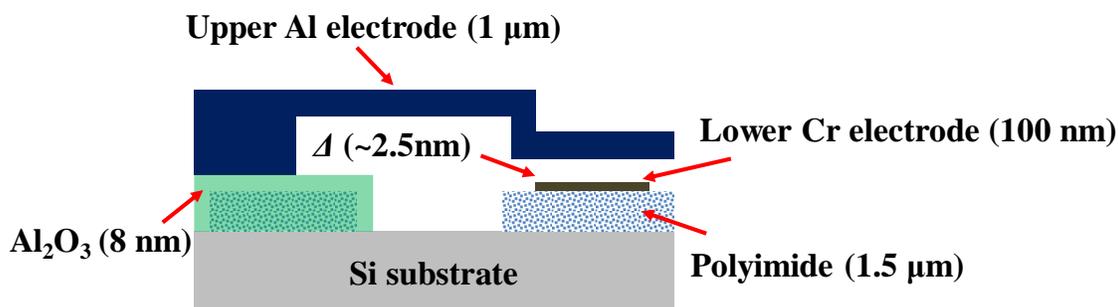

**Figure 5.2**: Schematic of the fabricated device.

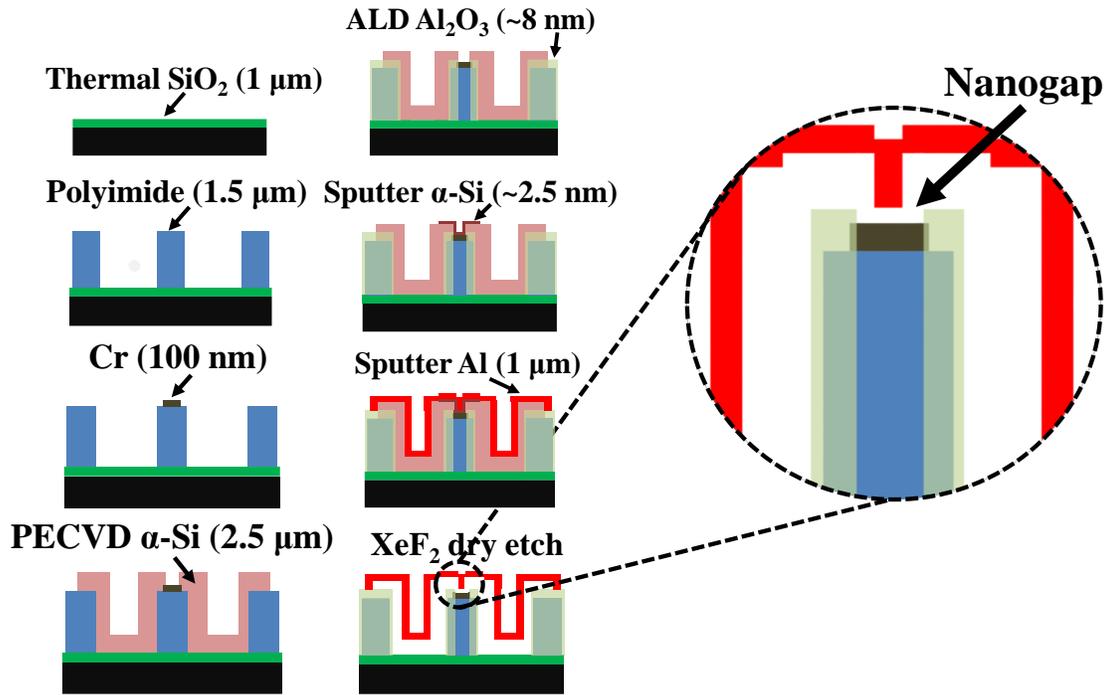

**Figure 5.3**: Simplified fabrication flow and zoomed in view of the nanogap junction.

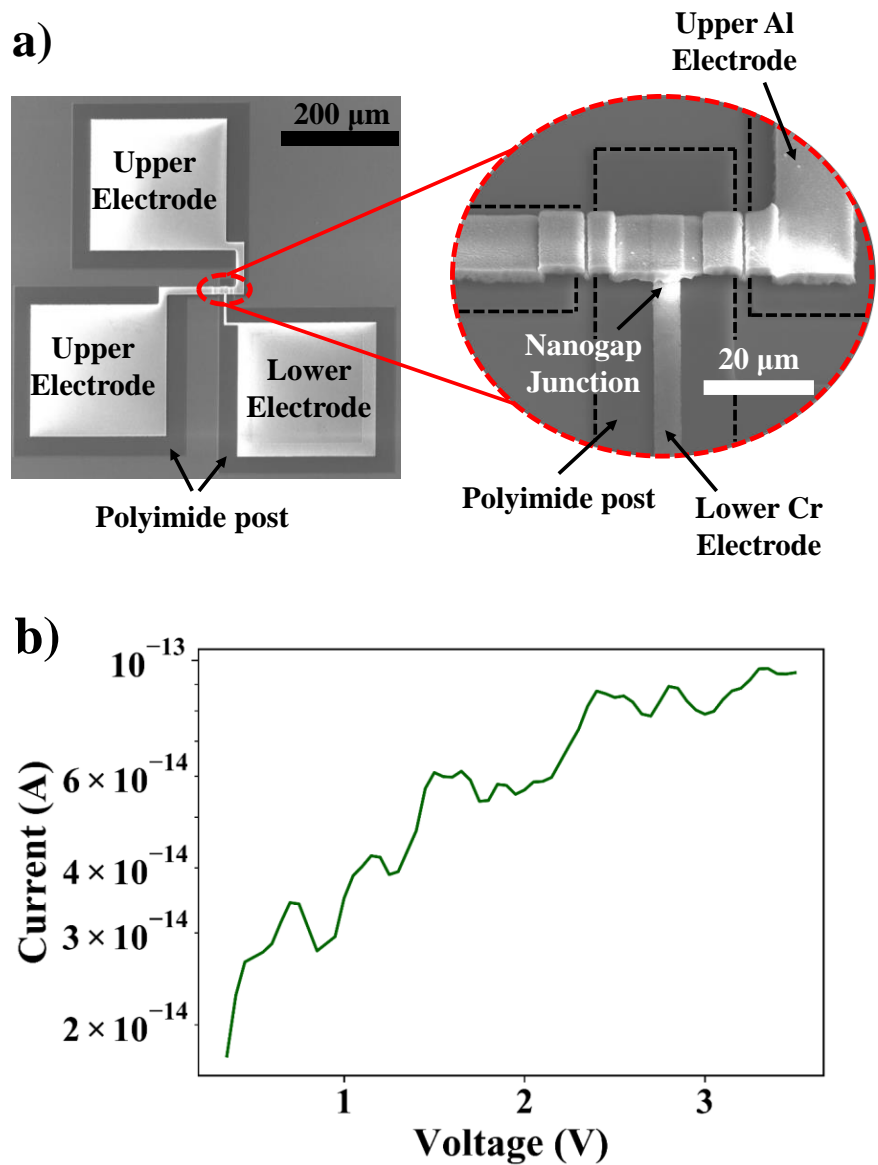

**Figure 5.4**: Device structure and electrical characterization. a) SEM images of fabricated device. b) I-V characteristics across the tunneling junction after sacrificial etching of the α-Si.

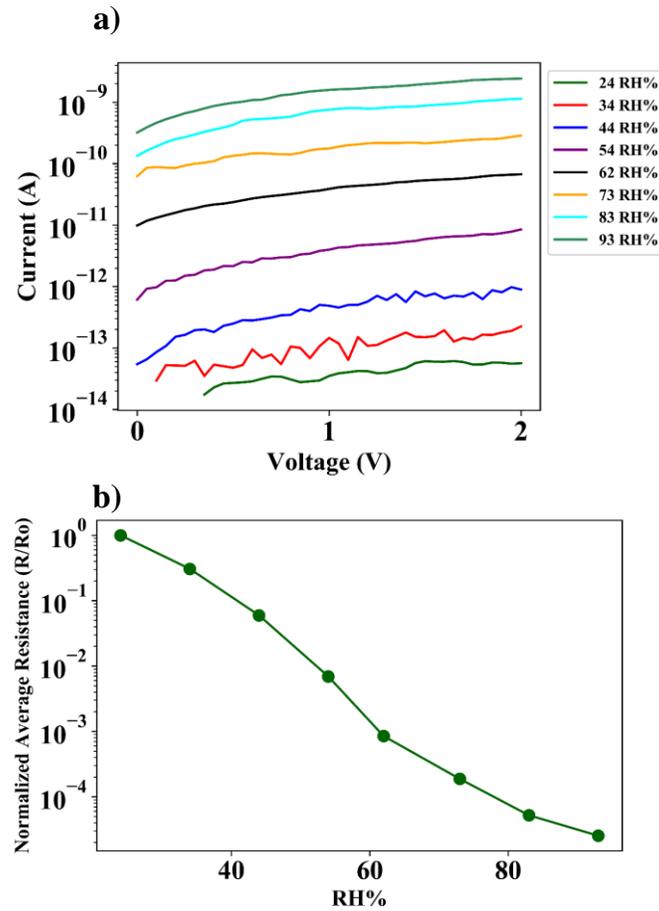

**Figure 5.5**: Device electrical response to change in ambient humidity. a) I-V characteristics of device. b) Normalized average junction resistance at various humidity levels in the test chamber.

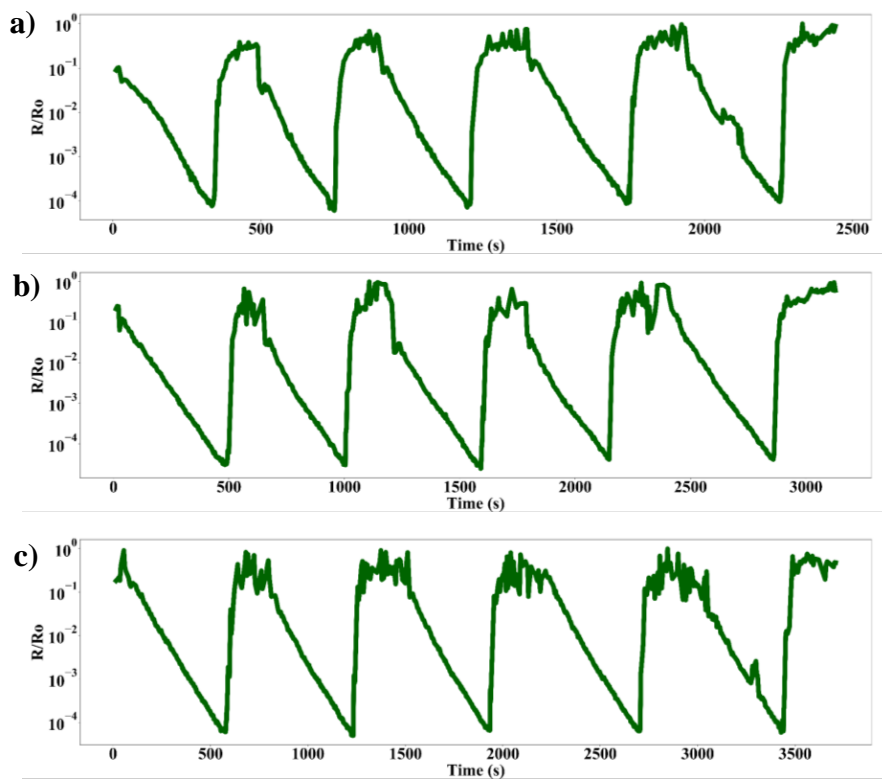

**Figure 5.6**: Repetitive cycles of water vapor exposure and removal from testing chamber for three different devices.

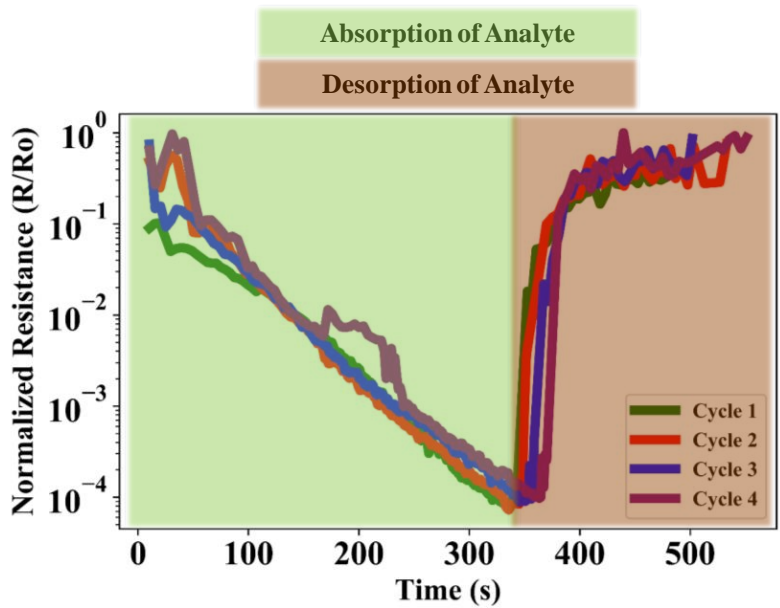

**Figure 5.7**: Comparison of sensor response for multiple cycles of moisture exposure and removal for the same device.

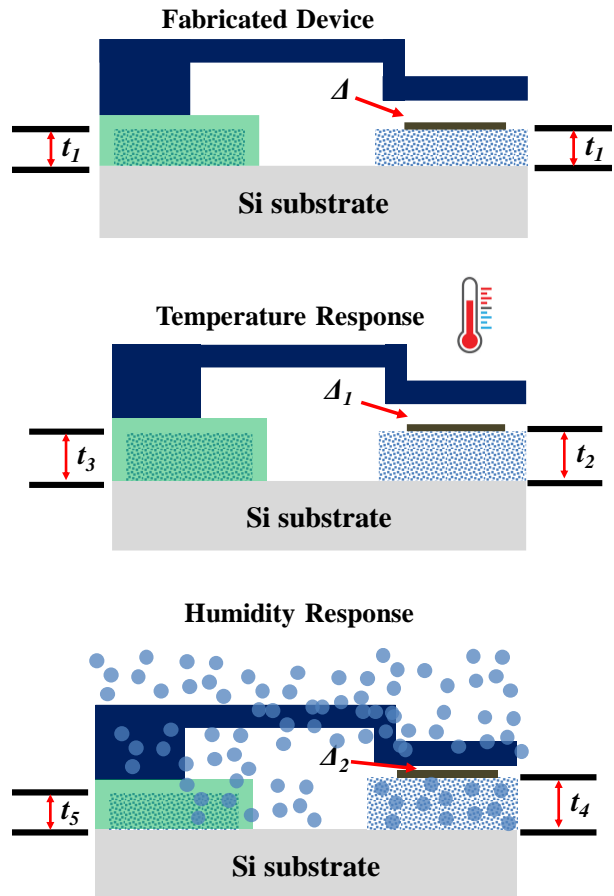

**Figure 5.8**: Schematic of nondifferential swelling design to achieve passive temperature compensation.

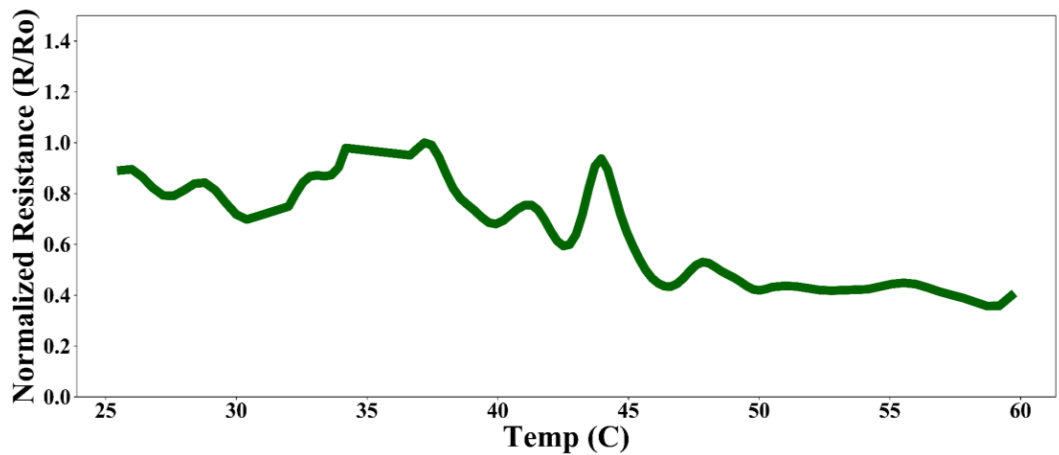

**Figure 5.9**: Temperature response of the device.

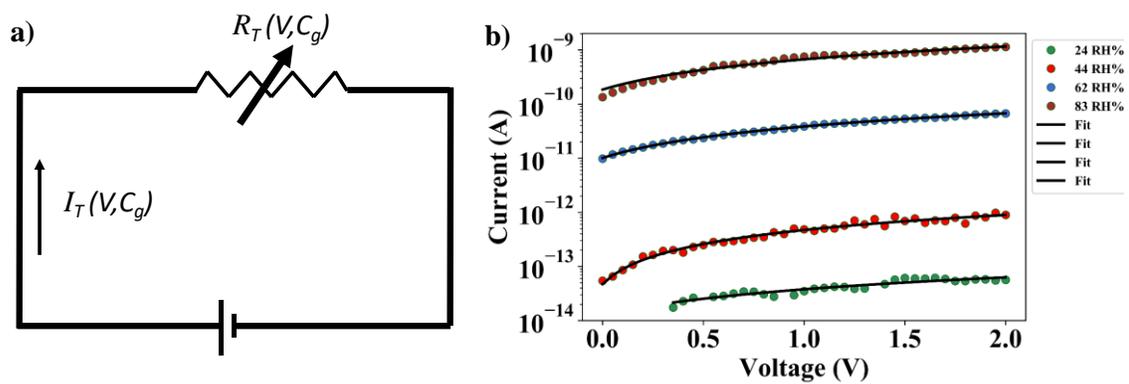

**Figure 5.10**: Mathematical modeling of tunneling sensor response. a) Equivalent electrical circuit of the sensor. b) Curve-fitted plots of the I-V measurements of the device after exposure to different levels of moisture content in the test chamber.

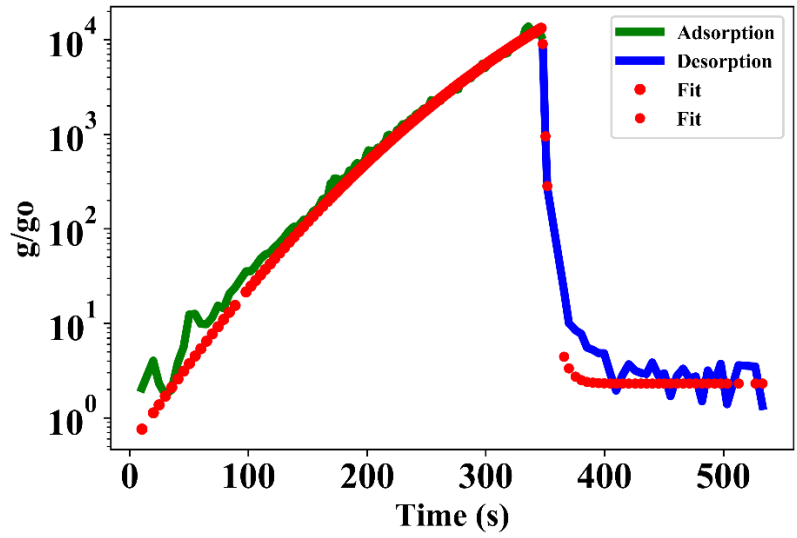

**Figure 5.11**: Normalized conductance response to one period of increasing exposure of humidity levels to nanogap device and subsequent removal, curve-fitted to Fick's law of diffusion and Polanyi-Wigner desorption model, respectively.

CHAPTER 6

A MILLIVOLT TRIGGERED MEMS PADDLE SWITCH

## 6.1 <u>Introduction</u>

Electrostatic MEMS switches exhibit high electrical isolation, abrupt switching characteristics, low on-resistance and they consume zero DC power; hence they have been extensively used in RF and biomedical devices [1]. However, the primary disadvantage of MEMS switches is that they require high voltages in the range of several tens of volts for efficient actuation. The high actuation pull-in voltage proves to be a severe deterring factor in their wide-spread use [2]. Commercially available switches operate at ~100V [3]. Recently several rib type cantilever switch devices with extended gate electrodes have been reported where lower actuation voltage has been achieved without reducing the stiffness [4]. This was achieved by optimizing the electrode design. In this article, authors used a certain bias voltage to detach the beam which was stuck due to stiction. Although in the last couple of years, we have witnessed novel methods of stiction removal, compatible to conventional MEMS processes [11,12]. However, in spite of the optimization, the pull-in voltage was still in the range of tens of volts, but the actuation voltage can be reduced further when the device is driven in a differential configuration. Recently the voltage requirement was reduced to 3 V using a see-saw structure that pulls-in the see-saw to one of its tilted states [1].

In this paper we extend this principle to produce electrostatic MEMS devices that can switch state using a few mV. These devices operate on the concept of force-cancelling differential actuation. Figure 6.1 shows a schematic of the device and its electrical equivalent circuit. It consists of a see-saw dual wing paddle suspended by two torsional springs. The device is electrostatically driven by four symmetric electrodes placed beneath it. The inner set of electrodes is used to produce a large but balanced electrostatic torque that is canceled by to the biasing electrode symmetry. To operate the device the two inner electrodes are set to a fixed positive bias $V_B$, the outer electrodes are set to ground and input voltage signal $V_{in}$, and the paddle is set to a negative bias $V_P$. Although the net torque produced by the inner electrodes is zero, the nonlinearity of electrostatic forces produces continuous spring softening with bias hence large motion amplification [7] and ultimately pull-in. The introduction of a very small imbalance caused by the input voltage or $V_{in}$ can thus trigger the pull-in instability. The low pull-in voltage and its zero DC power consumption thus make this device suitable as a very sensitive signal trigger that can be used to close a switch and wake up a larger electronic system.

In this paper we present the bi-stable theory of the device trigger and we calculate the voltage threshold needed to induce the device snap-in. We also fabricated a test device. Experimentally we measured the pull-in trigger voltage to be as low as 50 mV.

## 6.2 Motion Amplification and Switching Behavior

### 6.2.1 Analysis of the MEMS differential see-saw switch

The analysis of similar bi-stable structures can be found in [in many forms [5-10]. The simplest explanation is provided by recognizing that the paddle equilibrium is obtained by minimization of the energy functional:

$$U_T = U_M - U_{EL},$$ (1)

where $U_M$ is the mechanical energy stored in the torsional spring and $U_{EL}$ is the electrical energy of all of the device capacitors. For the time being we will initially neglect the influence of the relay contacts so the formulation includes electrical energy terms from two inner and two outer electrodes. Note that the sign of $U_E$ in the energy functional is negative, as forces emerging from the electrical energy terms are attractive. This is also consistent with the co-energy formulation [9]. Figure. 6.2 shows a schematic of the device in the flat and collapsed states. The inner electrode set forms two capacitors $C_{i,L}$ and $C_{i,R}$ on the left and right wings of the paddle. Two additional capacitors $C_{o,L}$ and $C_{o,R}$ are formed between the paddle and the outer electrodes. The corresponding mechanical and electrical energies as a function of the rotation angle $\theta$ are:

$$U_M = \frac{1}{2} \cdot k_\theta \cdot \theta^2 \, , \quad U_{EL} = U_i + U_o$$ (2)

$$U_i = \frac{1}{2} \cdot (C_{i,L}(\theta) + C_{i,R}(\theta)) \cdot (V_B - V_P)^2$$ (3)

$$U_o = \frac{1}{2} \cdot (C_{o,L}(\theta) \cdot V_P^2 + C_{o,R}(\theta) \cdot (V_{in} - V_P)^2),$$ (4)

where the parameters ξ, λ, and ß are

$$\zeta(\theta) = \frac{a}{g} \cdot \theta$$ (5)

$$\lambda(\theta) = \frac{(b+c)}{g} \cdot \theta$$ (6)

$$\beta(\theta) = \frac{b}{g} \cdot \theta,$$ (7)

where θ is defined as positive in the anticlockwise direction. Under normal conditions, the voltage $V_B$ is set to a positive value and $V_P$ is set to a negative value while $V_{in}$ is very small. Therefore the term $U_o$ can be simplified as:

$$U_o \approx \frac{1}{2} \cdot (C_{o,L}(\theta) + C_{o,R}(\theta)) \cdot V_P^2 + C_{o,R}(\theta) \cdot V_P \cdot V_{in}. \tag{8}$$

The torsional spring constant depends on the dimensions of the suspension beam. For a suspension with two rectangular beams of width $w_b$, height $t_b$, length $L_b$ and shear modulus G is:

$$k_\theta = \frac{2 \cdot G \cdot w_b \cdot t_b^3}{3 \cdot L_b} \cdot (1 - \frac{192}{\pi^5} \frac{t_b}{w_b} \tanh(\frac{\pi}{2} \frac{w_b}{t_b})). \tag{9}$$

The static behavior of the paddle and electrode system is determined by the dependence of the energy functional versus the deflection angle.

## 6.2.2 Stability analysis at zero input voltage

One may expect that at zero input the entire structure is completely symmetric and the paddle would stay horizontally at rest. The paddle can only rest at locations where the energy functional is a minimum. Some minima are located at the functional critical angles θc which satisfy:

$$\left. \frac{\partial U_T}{\partial \theta} \right|_{\theta_c} = 0. \tag{10}$$

The critical points can be stable or unstable. This is best observed when one plots the curve of $U_T$ versus angle as shown in Figure 6.3. Note that the energy functional as the potential difference between the inner electrodes and the plate increases can have either one stable and two unstable critical points or no stable critical points. In addition to its critical points one must consider energy wells and local minima formed by hard limits on the paddle travel. In the unstable region of the curve the paddle will actually pull-in and rest at either end of the maximum paddle deflection on either the left or the right edge. This thus represents a bi-stable state.

After much manipulation one can determine that the entire structure becomes unstable when:

$$V_{B,crit} \geq V_P + \sqrt{V_{crit}^2 - \frac{c(c^2 + 3bc + 3b^2)}{a^3} \cdot V_P^2} \,, \tag{11}$$

where

$$V_{crit} = \sqrt{\frac{3 \cdot g^3 \cdot k_\theta}{2 \cdot \varepsilon \cdot a^3 \cdot w}} \tag{12}$$

is the critical bias voltage on the electrodes for instability when $V_P = 0$. Note that if $V_P < 0$ then this critical voltage will be reduced. Note that as the bias is increased the range of stable deflection is reduced until it becomes zero at the edge of stability.

Now let's consider what happens when $V_{in} \neq 0$. The simplest way to understand this is to consider the additional deflection due to the torque imbalance provided by $V_{in}$. This additional torque is calculated from

$$\tau(V_{in}) \approx \frac{\partial U_o}{\partial \theta}\bigg|_{\theta=0} = -\frac{\varepsilon \cdot (b + c/2) \cdot c \cdot w_p}{g^2} \cdot V_P \cdot V_{in}. \tag{13}$$

The deflection is equal to this torque divided by the effective torsional spring constant. However the effect of the strong bias is to soften the overall spring constant of the system. In fact the pull-in instability with no input voltage is caused by the effective spring constant becoming zero. The effective spring constant and deflection angle are thus

$$k_{eff} = k_\theta \cdot (1 - \frac{(V_B - V_P)^2}{V_{crit}^2} - \frac{c(c^2 + 3bc + 3b^2)}{a^3} \cdot \frac{V_P^2}{V_{crit}^2}) = \frac{k_\theta}{M}$$
(14)

$$\theta(V_{in}) = -\frac{M \cdot \tau(V_{in})}{k_\theta}.$$
(15)

The deflection angle is thus magnified by the factor M>>1. The onset angle for pull-in is approximately

$$\| \theta_{pullin} \| = \sqrt{\frac{5 \cdot k_{eff}}{6 \cdot k_\theta}} \frac{g}{a} \cdot \left( \frac{(V_B - V_P)^2}{V_{crit}^2} + \frac{c(c^4 + 5c^3b + 10b^2c^2 + 10cb^3 + 5b^4)}{a^5} \cdot \frac{V_P^2}{V_{crit}^2} \right)^{-\frac{1}{2}}$$
(16)

Note that this angle approaches zero as $k_{eff}$ becomes zero. Therefore not only the angle is magnified but also the threshold for onset of pull-in is reduced. Because of these three combined effects, it is possible to induce pull-in with just a few mV of input voltage and the closure of a switch. We have microfabricated a simple test device as discussed below.

## 6.3 Device Fabrication

Figure 6.4 shows the test device fabrication process. The process starts by deposition of 0.1 μm of thermally grown oxide. Next we sputter 0.1 μm of chrome on the surface and perform lithography of the device electrodes. The electrodes are patterned using wet chrome etchant. In the next step we deposit 2 μm of sacrificial PECVD amorphous silicon. Next we perform lithography on the sacrificial silicon layer to pattern the stiffeners and the anchor of the paddle. After this, S1813 photoresist is spin-coated and patterned for lift-off. Next, 50nm of Cr and 100nm of Au is sputtered on the sample. After this, thick AZ 9260 photoresist is spin coated on the sample and lithography is done to create a mask for the electroplating process which is done next. The Cr-Au layer acts a seed layer for electroplating of Ni, which is deposited to a thickness of 4 μm. After this, lift-off is done by dipping the sample in acetone and putting it in an ultrasonic chamber. Finally, after lift-off patterning of the paddle, the sacrificial layer is etched away isotropically using $XeF_2$. After about ~7 hours of etching, the sacrificial layer is removed. Figure 6.5 shows a photograph of the test device. Each paddle wing measures 500 μm in length.

## 6.4 <u>Results and Discussion</u>

The see-saw paddle is biased as shown in Figure 6.2. The two inner electrodes are positively biased at VB. One of the outer electrodes is provided with a positive input voltage $V_{in}$ and the other outer electrode is kept at 0 V. The plate voltage $V_p$ is maintained at a negative voltage to reduce the stiffening of the paddle. The $V_{in}$ is applied from 0v to 500mv to observe snap-in. To electrically confirm the snap-in, capacitance measurement between the paddle and the underlying electrode is done using an Agilent 4284 Precision LCR meter. The snap-in is also verified using SEM imaging and subsequent profiling by the zygo optical profilometer.

On application of $V_{in}$ = 500 mV ($V_B$ = 40V and $V_P$ = -27V), $V_{in}$ = 100mV ($V_B$ = 54V and $V_P$=-27V) and $V_{in}$ = 50mV ($V_B$ = 60V and $V_P$ = -30V), snap-in of the paddle was observed. Snap-in was electrically confirmed by measuring capacitance of the paddle wing with input voltage before and after application of voltages. Figure 6.6 shows the resultant change in capacitance on snap-in. The SEM image of the paddle before and after snap-in is shown in Figure 6.7. The snap-in is confirmed with an optical profilometer as shown the inset of Figure 6.7 The paddle shows efficient switching operation at extremely low input voltages (50-500 mV) with zero DC power consumption.

## 6.5 <u>Application to Highly Sensitive Vapor Sensing</u>

Using the above established spring-constant softening technique, a photo-resist coated paddle structure was used for highly sensitive vapor sensing. Figure 6.8 shows the working principle of the device and the schematic of the polymer coated paddle switch.

As part of the application, we present a chemo-mechanical sensor with parametrically magnified deflection. The deflection can be sensed by measuring change in capacitance which does not require cumbersome optical testing. Our goal is the magnification of the adsorption-induced deflection by at least 10-fold resulting in snap instability. The deflection amplifying device consists of a micromachined paddle (a see-saw) supported by two torsion springs. One of the paddle wings is coated with a thin adsorptive polymer layer. The paddle is electrically grounded and its deflection is driven by two symmetrically placed electrodes beneath it spaced by gap g. The electrodes are connected to a DC bias voltage $V_B$. If there is no chemically induced force in the polymer and if the paddle is flat, the torque from the bias electrodes is the same on both sides of the paddle the net torque is zero and the default state of the paddle is horizontal at rest. When this device is exposed to adsorbed vapor, the polymer coating expands causing one of the paddle to warp and deflect downwards. This absorption induced deflection is magnified by a nonlinear gain mechanism which is provided by applying DC biases on the substrate electrodes. The nonlinearity of the electrostatic force on the paddle produces spring softening and parametric magnification of motion. If the biases are sufficiently high, pull-in snapping behavior is also observed.

### 6.5.1 Device fabrication

The device was fabricated using a similar method as given in Section 6.2. Figure 6.9 shows the simplified fabrication process Figure 6.10 shows optical and high-resolution SEM images of the fabricated device.

### 6.5.2 Preliminary results

The effect of bias voltage and deflection amplification due to nonlinear gain amplification was examined. Analytical calculations of the device geometry provided an estimation of $V_{crit}$ which was calculated to be ~10V. Hence a maximum bias voltage of 12V was chosen for experiments. In a first set of experiments the paddle wing deflection was measured under a profilometer in the absence of acetone vapors but in the presence of bias voltages. Negligible deflection was observed through which we can conclude that our device was fairly rigid and symmetrical. The same device was then exposed to acetone vapors at atmospheric pressure and a bias voltage of 12V. As shown in Figure 6.11. We observed a 12-fold increase in deflection in the presence of acetone vapors and bias voltage.

The paddle pull-in behavior was next observed by plotting the time resolved capacitance measurement of the paddle wing with resist patch. As the paddle deflects the capacitance value increases as given by (5). Two devices of same dimensions were exposed to acetone vapors under similar conditions. In one of the devices no DC bias was given while the other device was given a 12V DC bias. Figure 6.12 shows the time-resolved capacitance measured for the two devices, where the capacitance was measured with an Agilent 4284 LCR meter at an interval of 10 seconds. The device without bias shows only a small variation in capacitance value. The maximum capacitance value without bias corresponds to a deflection of ~150nm which is consistent with the observation in Figure 6.7. In the presence of a DC bias we observed a sudden jump in capacitance value signifying the occurrence of instability and pull-in. The maximum capacitance value observed (~150 fF) was approximately equal to the calculated value of capacitance at pull-in (155 fF).

## 6.6 <u>Conclusion</u>

This chapter proposed a solution to the most basic problem of MEMS switches: high actuation or pull-in voltage using a differential symmetrically biased torsional see-saw paddle. The electrically conductive paddle was symmetrically biased by applying voltage at its inner electrodes which lead to the paddle experiencing a bi-stable. The symmetric biasing softens the stiffness of the paddle which leads to a very low pull-in voltage. The fabricated demonstration device was provided an input voltage Vin of 50mV-500mV (for different values of biasing voltages) and snap-in was confirmed electrically (by measuring the capacitance between the paddle and the electrode with the LCR meter), by SEM imaging and by the optical profilometer. Snap-in has been successfully shown to take place on application of ultra-low mV voltages.

Using this concept of spring-softening, a chemo-mechanical sensor based on nonlinear parametric amplification of displacement of a see-saw paddle under exposure to analyte vapors. This sensor scheme does not require cumbersome detection instruments and works on zero-DC power utilizing a small DC-bias voltage to trigger instability induced displacement amplification. We have reported the displacement amplification and pull-in behavior of such a sensor using photoresist as polymer and acetone vapors as analyte. A 12 fold increase deflection amplitude was observed on application of 12 V DC bias.

## 6.7 <u>References</u>

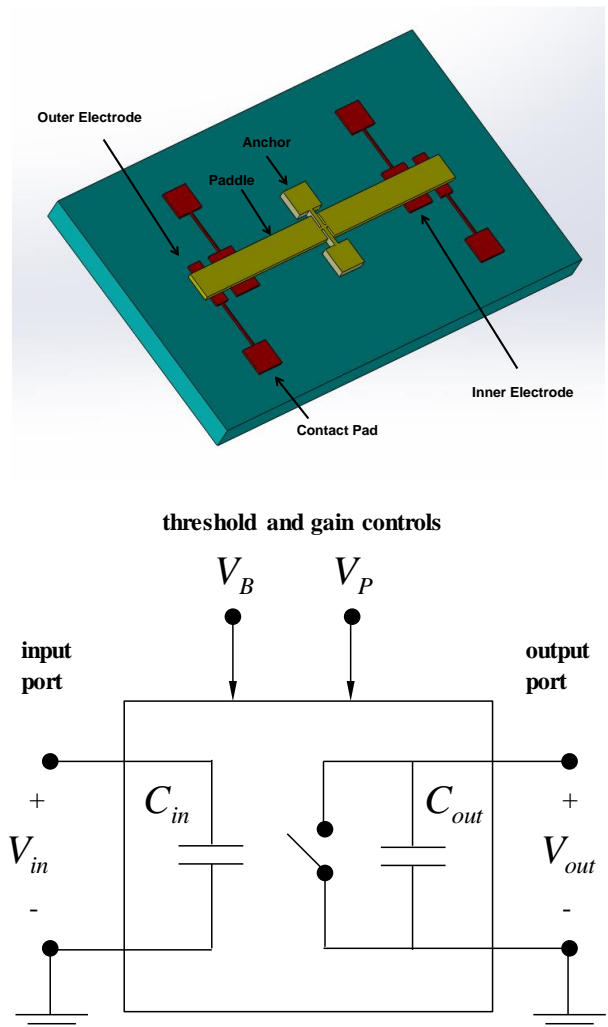

**Figure 6.1**: Schematic of the low pull-in voltage MEMS paddle switch and electrical

equivalent circuit.

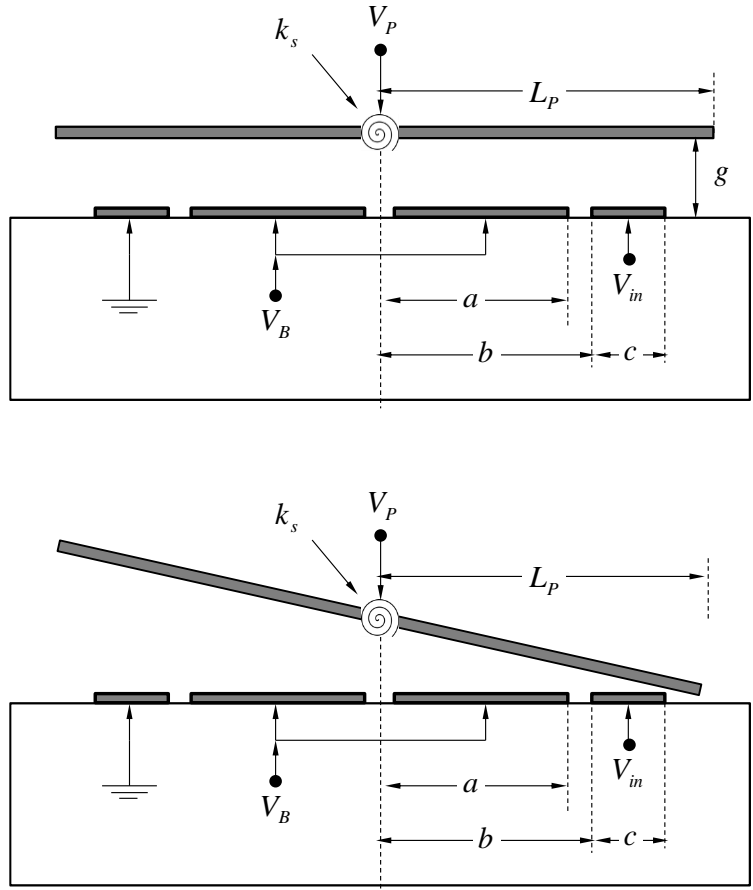

**Figure 6.2**: Biased see-saw paddle before and after snap-in.

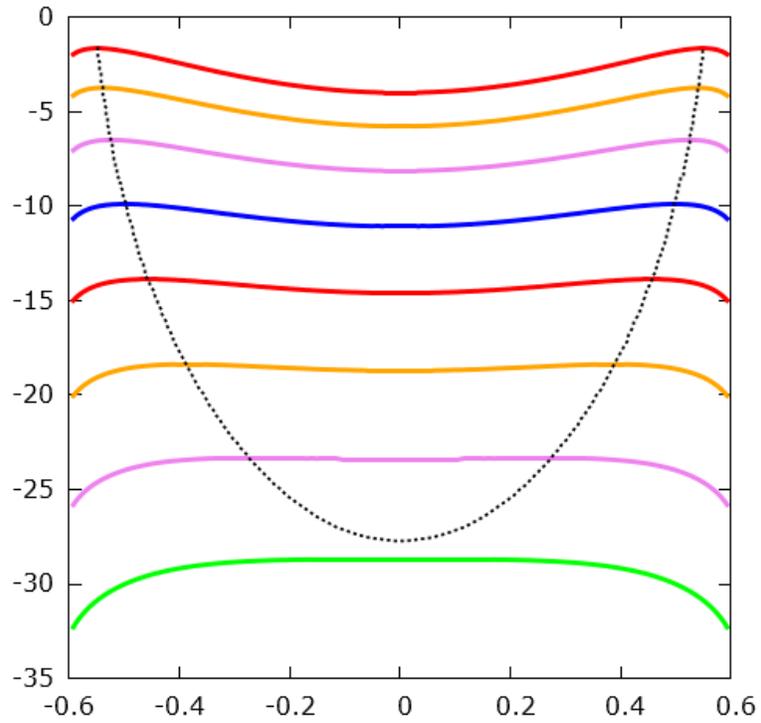

**Figure 6.3**: Example system energy for rotational voltage comparator under symmetric bias. For low VB the plate is horizontally balanced but will pull in if the deflection exceeds the unstable equilibria. For $V_B > V_{crit}$ there are no stable points.

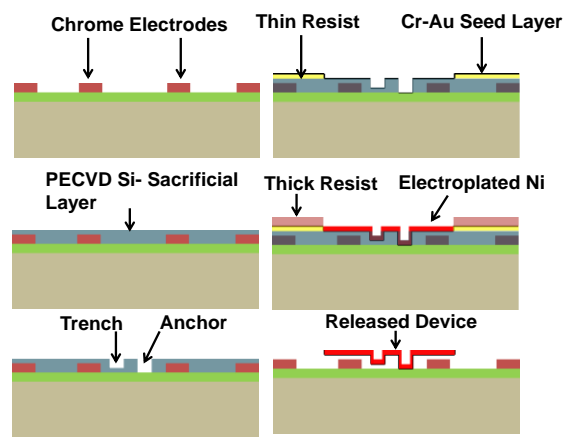

**Figure 6.4**: Device fabrication process.

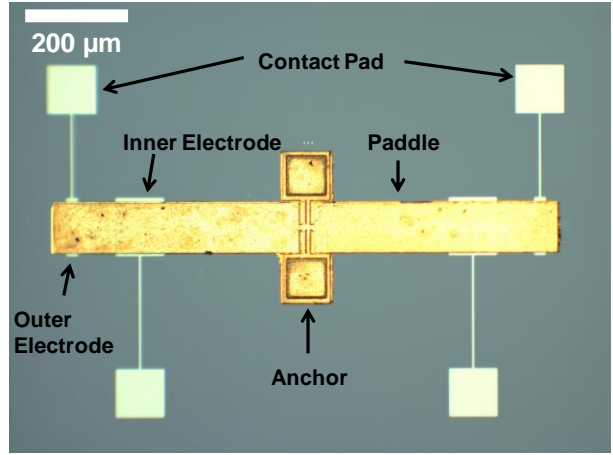

**Figure 6.5**: Optical image of MEMS see-saw paddle switch.

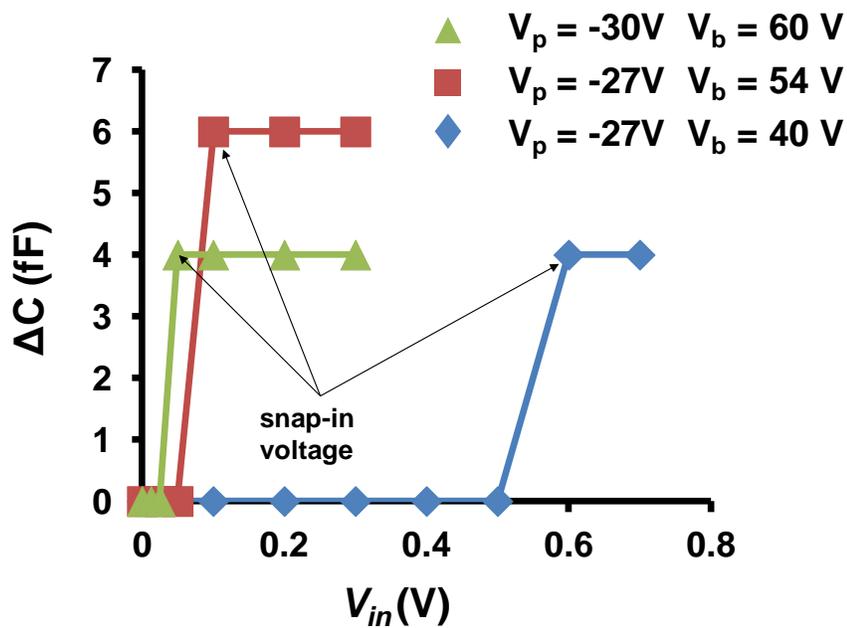

**Figure 6.6**: Change in paddle capacitance ($\Delta C$) vs $V_{in}$ measured at 10mV interval of applied Vin showing the pull-in transition at 50, 100 and 500 mV under different biasing DC voltages. Measurement is taken by switching to LCR meter at the end of each $V_{in}$ interval.

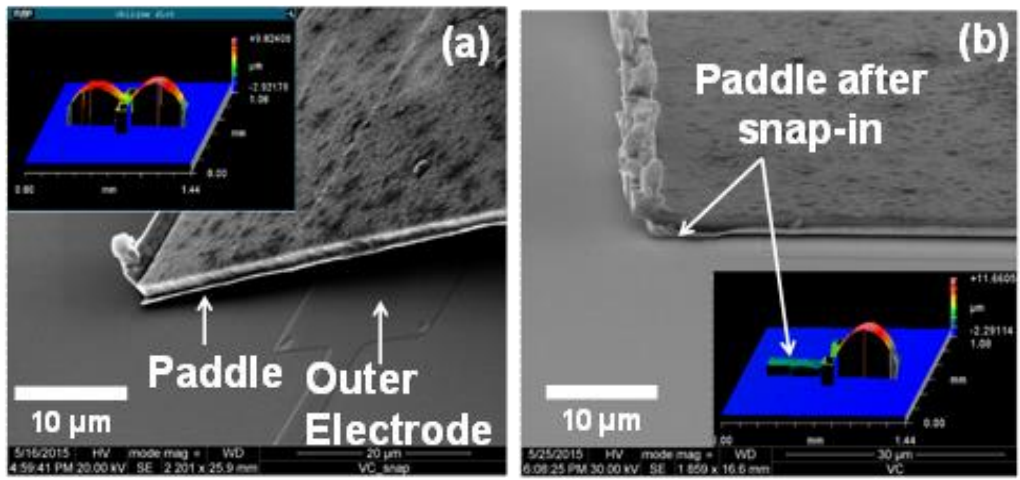

**Figure 6.7**: SEM of paddle wing before and after pull-in achieved with V$_{in}$.

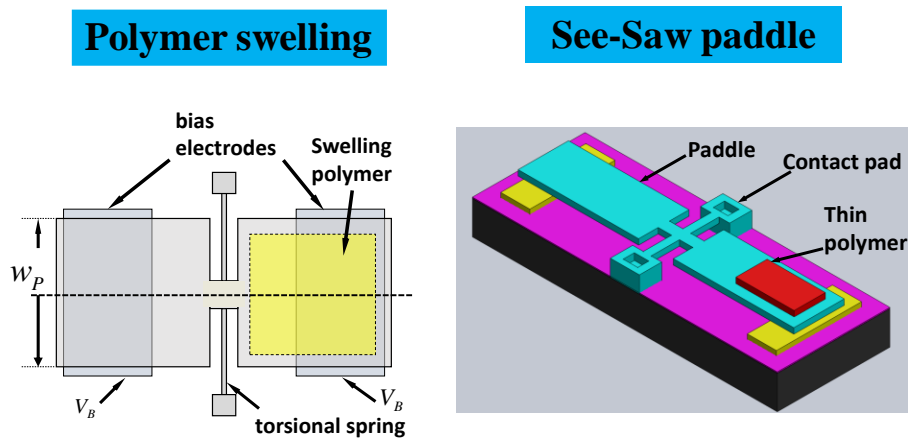

**Figure 6.8**: Schematic of vapor sensitive amplified chemo-mechanical sensor. One paddle wing is covered with a vapor sensitive polymer and a DC bias is applied to bottom electrodes.

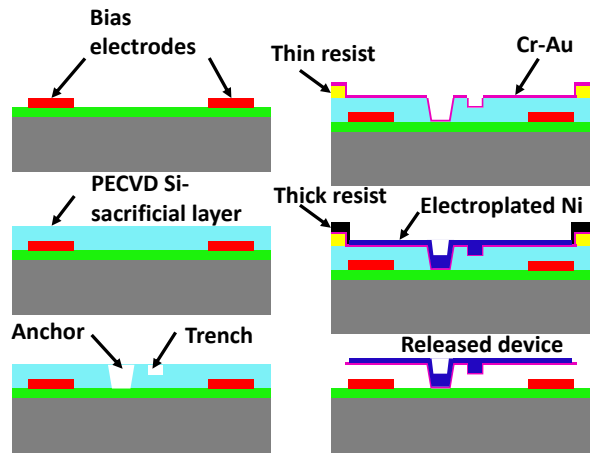

**Figure 6.9**: Simplified fabrication process.

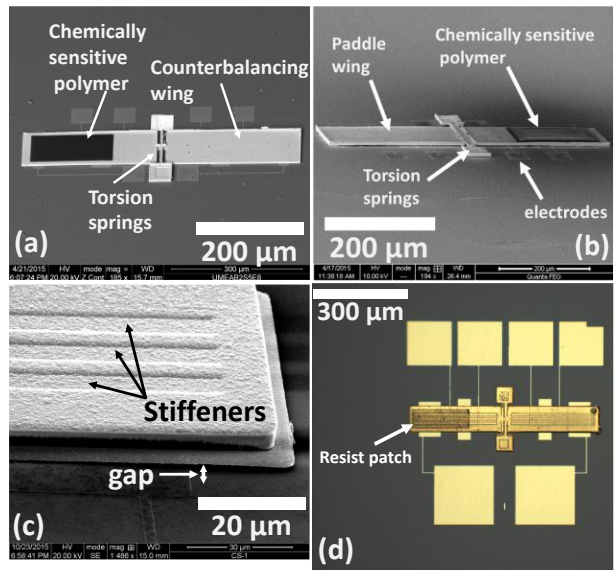

**Figure 6.10**: Optical and SEM images of the released vapor sensor device with a polymer patch on one paddle wing.

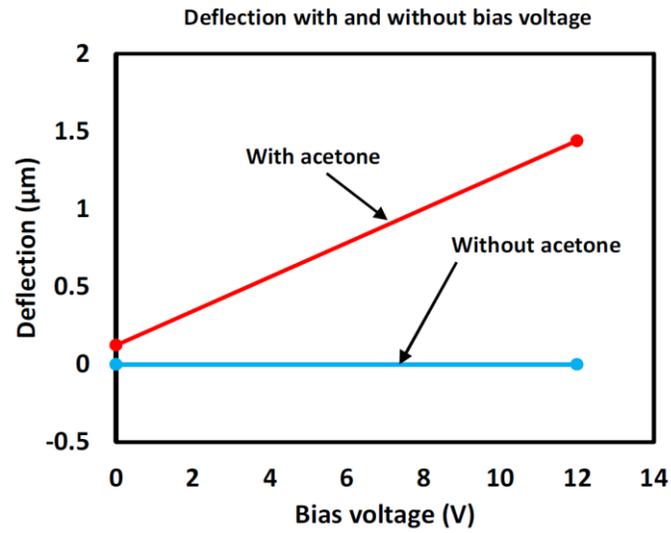

**Figure 6.11**: Effect of bias voltage resulting in a 12 fold deflection magnification in the presence of acetone vapors compared to the same device with zero bias.

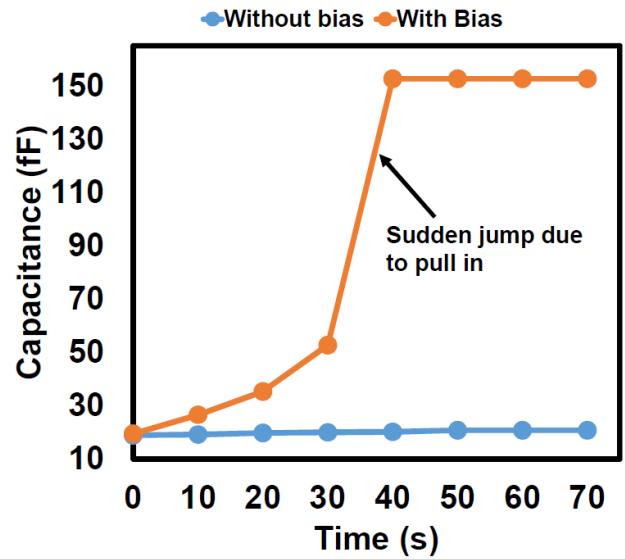

**Figure 6.12**: Time resolved capacitance measurement shows the pull-in behavior.

CHAPTER 7

CONCLUSIONS

The aim of this research was the development of a new class of chemiresistors which consume extremely low power during device operation, are highly sensitive and display low cross-sensitivity towards commonly found VOCs and gases. In accordance with these aims, we fabricated two new types of gas-sensors. The first was a nanogap device, which was designed such that the functionalized device was exposed to the analyte, the sensor would trap the intended analyte molecules within the nanogap and the junction resistance would reduce by several orders of magnitude while only consuming 15 pW of DC power during "stand-by" mode. The second type of sensor was a quantum tunneling hygrometer which displayed several orders of resistance change when exposed to water vapor molecules. The humidity sensor consumed 0.4 pW of power during "stand-by" mode. Both the devices were batch-fabricated and are compatible with existing CMOS technology. They are therefore suitable for IoT based applications and low power sensing. The significant contributions of this work are listed below.

- Nanogap electrodes with spacer thicknesses as low as ~4.0 nm were fabricated and we performed extensive electrical characterization of the fabricated devices across a 4-inch wafer. A novel use of α-Si was also demonstrated as a high resistance adhesive layer for gold.

- Using standard optical techniques, extensive thickness uniformity characterization of the deposited ultra-thin films and process repeatability measurements were performed.

- Tunneling current measurements were performed on devices across the wafer and nonuniformities in I-V characteristics were investigated. Transition Voltage Spectroscopy method was used to determine the barrier potential of the spacer film.

- Breakdown measurements were also performed to determine the maximum operating voltage of these devices. Temperature response of the device was also monitored by measuring the I-V characteristics of the device at different substrate temperatures.

- The fabricated devices were functionalized with a fully conjugated ter-phenyl linker molecule, (4-((4-((4 mercaptophenyl) ethynyl) phenyl) ethynyl) benzoic acid) for electrostatic capture of our target gas - cadaverine.

- We demonstrated ultra-low power resistance switching in batch-fabricated nanogap junctions upon detection of target analyte - cadaverine. The stand-by power consumption was measured to be less than 15.0 pW and the $R_{OFF}/R_{ON}$ ratio was more than eight orders of magnitude when exposed to ~80 ppm of cadaverine.

- A phenomenological electrical model of the device is also presented in good agreement with experimental observations.

- Cross-sensitivity of the gas sensor was tested by exposing the device to a variety of the commonly found VOCs and other atmospheric gases. The experiments revealed a highly selective sensor action against most of these analytes.

- These batch-fabricated sensors consume ultra-low power and demonstrate high selectivity; therefore, they can be suitable candidates for sensor applications in power-critical IoT applications and low power sensing.

- The design, fabrication, electrical characterization and working of a temperature compensated tunneling humidity sensor was presented. Sensor response showed a completely reversible reduction of junction resistance of ~four orders of magnitude with a standby DC power consumption of ~0.4 pW.

- Passive temperature compensation was also achieved using a nondifferential swelling design for the polymer patches. Temperature response showed a reduction of ~2.5 times when the device was exposed to a temperature sweep of 25°C – 60°C, which is 0.0025% of the maximum sensor output when exposed to rising levels of humidity.

- A mathematical model and equivalent electrical circuit was also presented to accurately describe the current conduction mechanism responsible for sensor action. Finally, sensor response dynamics was also investigated and established sorption analytical models were used to describe the time dependent sensor action.

The fabricated devices represent a new type of ultra-low power chemiresistive sensors. Since they are first-generation devices, they demand improvements before they can be implemented commercially. Some of these are as follows:

- In order to improve the uniformity of the deposited thin films, advanced

deposition techniques such as Pulsed Laser Deposition can be used during the fabrication process. This would also ensure a more selective sensing action based on the length of the analyte molecule.

- Since the molecules are captured in the nanogap present along the perimeter of the upper electrode in the nanogap device, an Auxetic-Fractal electrode design would ensure a higher number of capture sites whilst keeping the overlap area and by extension, foot-print of the device the same.

- Although we used a phenomenological electrical model to accurately describe sensor action which was in good agreement with the experimental data, the model was unsuccessful in providing quantitative measurements of energy levels of the individual molecules forming a bridge across the junction. In order to tackle this issue one must look towards using computational chemistry methods such as NEGF and DFT formulism to better quantify important junction characteristics of the molecular channel.

- Sensor recovery was a major issue that we faced during our experiments. This requires further characterization and assessment of sensor action and target molecules to enable enhanced recovery of the nanogap sensors.

- Although we did provide passive temperature compensation for our humidity sensor, experiments still showed a nonzero temperature response. Additionally, the device does not compensate for temperature-dependent polymer absorption behavior. Both these aspects can be addressed by improved designs to the humidity sensor.